\newif\ifsubmode
\submodefalse

\newif\ifprintfig
\printfigtrue

\ifsubmode
  \documentstyle[12pt,aasms4,epsf]{article}
  \received{}
  \accepted{}
  \journalid{}{}
  \articleid{}{}
\else
  \documentstyle[11pt,aaspp4,epsf]{article}
  \slugcomment{}
\fi

\lefthead{van den Bosch, Lewis, Lake \& Stadel}
\righthead{Orbital eccentricities of dark halos}

\newcommand{\etal}{{et al.~}}

\newcommand{\lta}{\lesssim}
\newcommand{\gta}{\gtrsim}
\newcommand{\lte}{\leq}

\newcommand{\kms}{\>{\rm km}\,{\rm s}^{-1}}

\newcommand{\Msun}{\>{\rm M_{\odot}}}

\begin{document}

\title{Substructure in Dark Halos: Orbital Eccentricities and
       Dynamical Friction}

\author{Frank C. van den Bosch\altaffilmark{1,3,5}, 
        Geraint F. Lewis\altaffilmark{1,2,4}, 
        George Lake\altaffilmark{1}, 
        Joachim Stadel\altaffilmark{1}}

\bigskip

\affil{$^{1}$ Department of Astronomy, University of Washington, Seattle, 
       WA 98195, USA}
\affil{$^{2}$ Department of Physics and Astronomy, University of Victoria, 
       Victoria, B.C., Canada} 


\altaffiltext{3}{Hubble Fellow}
\altaffiltext{4}{Fellow of the Pacific Institute for Mathematical 
                 Sciences 1998-1999}
\altaffiltext{5}{e-mail: vdbosch@hermes.astro.washington.edu}

\setcounter{footnote}{5}


\ifsubmode\else
\clearpage\fi


\ifsubmode\else
\baselineskip=14pt
\fi


\begin{abstract}
  The virialized regions of  galaxies and clusters contain significant
  amounts  of substructure;  clusters have  hundreds   to thousands of
  galaxies,  and satellite systems  and   globular clusters orbit  the
  halos of individual    galaxies.  These orbits  can decay   owing to
  dynamical friction.  Depending on their orbits and their masses, the
  substructures either merge, are disrupted  or survive to the present
  day.
    
  We examine the distributions of eccentricities of orbits within mass
  distributions like  those  we see  for  galaxies  and clusters.    A
  comprehensive understanding of these orbital properties is essential
  to  calculate the  rates of   physical   processes relevant  to  the
  formation and evolution of galaxies and clusters.

  We  derive the  orbital  eccentricity distributions  for a number of
  spherical potentials.   These distributions depend   strongly on the
  velocity    anisotropy, but only   slightly    on the  shape of  the
  potential.    The eccentricity distributions   in   the case  of  an
  isotropic distribution function   are strongly  skewed towards  high
  eccentricities,  with   a  median value   of  typically  $\sim 0.6$,
  corresponding to an apo- to pericenter ratio of $4.0$.
   
  We also present high resolution $N$-body  simulations of the orbital
  decay of satellite systems   on  eccentric orbits in  an  isothermal
  halo.  The dynamical friction timescales are  found to decrease with
  increasing orbital eccentricity  due to  the dominating deceleration
  at   the   orbit's  pericenter.   The   orbital   eccentricity stays
  remarkably constant throughout the  decay; although the eccentricity
  decreases near pericenter,  it increases again  near apocenter, such
  that there is no net circularization.

  We briefly    discuss   several  applications  for     our   derived
  distributions of orbital   eccentricities  and the  resulting  decay
  rates  from    dynamical  friction.   We   compare  the  theoretical
  eccentricity distributions to    those  of  globular  clusters   and
  galactic satellites for which all  six phase-space coordinates  (and
  therewith their  orbits) have  been  determined.  We  find  that the
  globular clusters are consistent with  a close to isotropic velocity
  distribution, and they show  large orbital eccentricities because of
  this (not  in spite of  this, as has been  previously asserted).  In
  addition, we  find that the  limited data on  the galactic system of
  satellites  appears   to     be different  and      warrants further
  investigation as a clue to the formation  and evolution of our Milky
  Way and its halo substructure.
\end{abstract}


\keywords{cosmology: dark matter ---
          galaxies: kinematics and dynamics ---
          galaxies: structure ---
          galaxies: interactions ---
          globular clusters ---
          stellar dynamics ---
          numerical methods.}

\clearpage


\section{Introduction}
\label{sec:intro}

In  the  standard cosmological  model  galaxies and  clusters  form by
hierarchical  clustering  and merging   of small density perturbations
that grow by gravitational  instability.  In this standard picture the
mass of the Universe is dominated by dissipationless dark matter which
collapses to  form dark halos, inside of  which  the luminous galaxies
form. It was once assumed that the previous generation of substructure
was erased at  each  level of  the   hierarchy (White \&  Rees  1978). 
However, high resolution $N$-body simulations have recently shown that
some  substructure is preserved  at  all levels  (Moore, Katz \&  Lake
1996; Klypin \etal 1997; Brainerd,  Goldberg \& Villumsen 1998;  Moore
\etal 1998; Ghigna \etal  1998).  This is consistent with observations
which reveal substructure in a  variety of systems: globular  clusters
within  galaxies, distant  satellites  and globulars  in  the halos of
galaxies, and galaxies within  clusters.  This substructure evolves as
it  is subjected  to the forces   that  try to dissolve  it: dynamical
friction, tides from the central objects and impulsive collisions with
other substructure.  The timescales for survival and the nature of the
dissolution guide  our understanding of  the formation processes, most
of which depend  on the nature of  the orbits.  Tides strip satellites
on elongated  orbits.  An elongated  orbit and  a circular one clearly
evolve differently owing to dynamical friction, especially if there is
a disk involved.  The disk heating similarly  depends on the orbits of
the satellites.  The nature of the  orbits also effects the nature and
persistence of tidal streams at breakup,  and the mutual collisions of
structures as considered by galaxy harassment.

Since the clustering and  merging of halo substructures  is one of the
cornerstones of  the   hierarchical structure   formation  scenario, a
comprehensive   understanding  of  their   orbital properties   is  of
invaluable  importance  when seeking  to  understand the formation and
evolution of  structure in   the Universe.   We  have found  that  the
properties of  orbits within   spherical,  fully relaxed  systems  has
received  little attention  and is often   misrepresented.  Hence, the
first  goal of this paper  is to derive a statistical characterization
of  the   orbits in   potential/density  distributions   that describe
galaxies within clusters, globular  clusters within galaxies, etc.  We
find that orbits are far more  elongated than typically characterized. 
Recently,  Ghigna   \etal   (1998)   used  high  resolution   $N$-body
simulations to investigate    the orbital properties of   halos within
clusters.  They  found   that orbits of    the subhalos  are  strongly
elongated with a median  apo-to-pericenter ratio of approximately $6$. 
We  compare our results on  equilibrium  spherical potentials with the
distribution of   orbits in their  cluster that   was  simulated in  a
cosmological context, and show that the  orbital eccentricities of the
subhalos are consistent with    an isothermal halo  that is   close to
isotropic.

In the  second  part of this   paper, we use high  resolution $N$-body
simulations to  calculate the dynamical  friction on eccentric orbits. 
Past studies  have compared numerical simulations with Chandrasekhar's
dynamical  friction formula (e.g.,   White 1978, 1983;  Tremaine 1976,
1981; Lin \& Tremaine 1983; Tremaine \& Weinberg 1984; Bontekoe \& van
Albada 1987; Zaritsky \& White 1988; Hernquist \& Weinberg 1989; Cora,
Muzzio \& Vergne 1997).  Comprehensive overviews of these studies with
discussions regarding   the  local versus global  nature  of dynamical
friction can  be found in   Zaritsky \& White   (1988) and Cora  \etal
(1997).  Most  of the studies  followed the decay  of circular or only
slightly eccentric  orbits.  The two   exceptions are Bontekoe  \& van
Albada (1987)  and Cora \etal  (1997). The former examined the orbital
decay of a `grazing  encounter' of a  satellite on an elliptical orbit
that  grazes a  larger  galaxy  at   its  pericenter.  In  this  case,
dynamical  friction occurs  only  near   pericenter.  The  pericentric
radius  remains nearly fixed  with  significant circularization of the
orbit in just  a few  dynamical times  (see  also Colpi (1998) for  an
analytical treatment based  on linear  response theory).   Cora  \etal
followed satellites on  eccentric orbits that were completely embedded
in a dark halo,  but they didn't  discuss the dependence of decay time
on  orbital  eccentricity or the change   in orbital eccentricity with
time.  We use fully self-consistent $N$-body simulations with $50,000$
halo particles to  calculate dynamical friction  on  eccentric orbits. 
We show that the timescale for dynamical friction  is shorter for more
eccentric orbits, but that the  dependence is considerably weaker than
claimed  previously by  Lacey \&  Cole (1993) based   on an analytical
integration of  Chandrasekhar's  formula. In  addition, we  show that,
contrary  to common  belief,  dynamical  friction   does not lead   to
circularization.  All in all, dynamical friction  only leads to a very
moderate change  in  the distribution  of  orbital eccentricities over
time.

In Section~\ref{sec:orbecc} we   derive the  distributions  of orbital
eccentricities for a  number of  spherical densities/potentials  using
both  analytical  and  numerical  methods.  Section~\ref{sec:friction}
describes our $N$-body  simulations of dynamical friction on eccentric
orbits in an isothermal halo.   In Section~\ref{sec:applic} we discuss
a number of applications. Our results and conclusions are presented in
Section~\ref{sec:conc}.

\section{Orbital eccentricities}
\label{sec:orbecc}

\subsection{The singular isothermal sphere}
\label{sec:sing}

The flat rotation curves  observed  for spiral galaxies  suggest  that
their dark halos have density profiles that are not too different from
isothermal.   Hence, we  start our  investigation   with the  singular
isothermal  sphere, whose potential,  $\Phi$, and density, $\rho$, are
given by
\begin{equation}
\label{potdens}
\Phi(r) = V_c^2 \; {\rm ln}(r),\;\;\;\;\;\;\;\;\;\;\;\;\;
\rho(r) = {V_c^2 \over 4 \pi G r^2}.
\end{equation}
Here $V_c$ is the circular velocity, which is constant with radius.

\subsubsection{Analytical method}
\label{sec:analyt}

For  non-Keplerian potentials, in   which  the orbits are not   simple
ellipses, it is customary  to define a generalized orbital eccentricity
\footnote{Throughout this paper we    use $e$ to denote   the  orbital
  eccentricity, ${\rm e}$ refers to the base of the natural logarithm,
  and $E$ refers to the energy.}, $e$, as:
\begin{equation}
\label{eccen}
e = {r_{+} - r_{-} \over r_{+} + r_{-}}.
\end{equation}
Here $r_{-}$ and  $r_{+}$  are the  peri- and apocenter  respectively. 
For an orbit  with energy $E$ and  angular momentum $L$ in a spherical
potential, $r_{-}$ and $r_{+}$ are the roots for $r$ of
\begin{equation}
\label{rooteq}
{1 \over r^2} + {2[\Phi(r) - E] \over L^2} = 0,
\end{equation}
(Binney \&  Tremaine  1987).   For  each energy, the   maximum angular
momentum  is, for  a singular  isothermal sphere,  given by  $L_c(E) =
r_c(E) V_c$. Here  $r_c(E)$ is the radius of  the circular orbit  with
energy $E$ and is given by
\begin{equation}
\label{circrad}
r_c(E) = \exp\left[{{E - V_c^2/2} \over V_c^2}\right].
\end{equation}
Upon writing $L  = \eta  L_c(E)$ ($0  \lte \eta  \lte 1$)\footnote{The
  quantity $\eta$ is  generally called the  orbital circularity} , one
can rewrite equation (\ref{rooteq})  for a singular isothermal  sphere
such that  the apo- and   pericenter are given by the   roots for $x =
r/r_c$ of
\begin{equation}
\label{rootred}
{1 \over x^2} + {2 \over \eta^2} {\rm ln}(x) - {1 \over \eta^2} = 0.
\end{equation}
As might be expected in this  scale free case, the ratio $r_{+}/r_{-}$
depends only on the  orbital circularity $\eta$  and is independent of
energy.  This dependence is shown in Figure~\ref{fig:eps}.
 
\placefigure{fig:eps}

At a certain radius, the average of any quantity $S$ is determined by
weighting it by the distribution function (hereafter DF) and
integrating over phase space. For the singular isothermal sphere this
yields
\begin{equation}
\label{avs1}
\overline{S}(r) = {4 \pi \over r^2 \rho(r)}
\int\limits_{\Phi(r)}^{\infty} dE \int\limits_0^{r\sqrt{2(E-\Phi)}}
\; f(E,L) \; S(E,L) {L \; dL \over \sqrt{2(E-\Phi) - L^2/r^2}}.
\end{equation}
In what follows we consider the family of quasi-separable DFs
\begin{equation}
\label{quasidf}
f(E,L) = g(E) \, h_a(\eta).
\end{equation}
This approach makes the solution of equation~(\ref{avs1}) analytically
tractable.  The general  properties  of spherical  galaxies  with this
family  of DFs are  discussed in  detail by  Gerhard (1991,  hereafter
G91).    We  adopt   a simple  parameterization      for the  function
$h_a(\eta)$:
\begin{equation}
\label{anisotropy}
h_a(\eta) = \left\{ \begin{array}{lll}
            \tanh \bigl( {\eta \over a} \bigr) / 
            \tanh \bigl( {1 \over a} \bigr) & \mbox{$a > 0$}\\
            1 & \mbox{$a = 0$}\\
            \tanh \bigl( { 1-\eta \over a} \bigr) / 
            \tanh \bigl( {1 \over a} \bigr) & \mbox{$a < 0$}
                   \end{array}
            \right.
\end{equation}
For $a=0$, the DF is isotropic. Radially  anisotropic models have $a <
0$, whereas positive  $a$ correspond to  tangential anisotropy.  For a
quasi-separable DF    of   the  form~(\ref{quasidf}),  and  for    the
eccentricity $e$ which  depends on $\eta$  only, equation (\ref{avs1})
yields
\begin{equation}
\label{avs2}
\overline{e}(r) = {4 \pi \over r \rho(r)}
\int\limits_{\Phi(r)}^{\infty} dE \; g(E) \; L_c(E)
\int\limits_0^{\eta_{\rm max}} h_a(\eta) \; e(\eta)
{\eta \; d\eta \over \sqrt{\eta_{\rm max}^2 - \eta^2}},
\end{equation}
where 
\begin{equation}
\label{etamax}
\eta_{\rm max} = {r \sqrt{2 (E - \Phi)} \over L_c(E)}.
\end{equation}
For a singular   isothermal sphere,  G91  has shown   that the  energy
dependence of the DF is given by
\begin{equation}
\label{df}
g(E) = { {\rm e} \over {16 \, \pi^2 \, G \, V_c \, u_H} } 
\exp\left[-{2 E \over V_c^2}\right],
\end{equation}
where
\begin{equation}
\label{uh}
u_H = \int\limits_{0}^{\infty} du \; {\rm e}^{-u}
\int\limits_{0}^{\eta_{\rm max}} h_a(\eta) \; {\eta \; d\eta \over
\sqrt{\eta_{\rm max}^2 - \eta^2}}. 
\end{equation}
Here $\eta_{\rm max}$ depends on $u$ only and is given by
\begin{equation}
\label{etamax_u}
\eta_{\rm max} = \sqrt{2 {\rm e}} \; \sqrt{u} \; {\rm e}^{-u}.
\end{equation}

Substitution of (\ref{df}) and (\ref{uh}) in (\ref{avs2}) yields (upon
substituting $u = (E - \Phi)/V_c^2$)
\begin{equation}
\label{avs3}
\overline{e}(r) = \overline{e} = {1 \over u_H}
\int\limits_{0}^{\infty} du \; {\rm e}^{-u} \int\limits_0^{\eta_{\rm
    max}} h_a(\eta) \; e(\eta) {\eta \; d\eta \over \sqrt{\eta_{\rm
      max}^2 - \eta^2}}.
\end{equation}
Note that,  due to the scale-free nature  of the problem, this average
is independent of radius.

\placefigure{fig:integral}

We  have   numerically  solved  this  integral  as  a function  of the
anisotropy     parameter    $a$.    The  results     are     shown  in
Figure~\ref{fig:integral}.   As  expected,  the  average  eccentricity
decreases going from   radial to tangential anisotropy.   The  average
orbital eccentricity  of an  isotropic, singular isothermal  sphere is
$\overline{e} =  0.55$.      An orbit  with  this eccentricity     has
$r_{+}/r_{-} = 3.5$.

\subsubsection{Numerical method}
\label{sec:numer}

To examine the actual distribution of eccentricities, rather than just
calculate  moments  of  the distribution,  we Monte  Carlo  sample the
quasi-separable   DF  of  equation~(\ref{quasidf})    for orbits  in a
singular isothermal potential and  calculate their eccentricities.  We
provide a detailed description of this method in the Appendix.

\placefigure{fig:anis}

The normalized  distribution functions  of orbital  eccentricities for
three  different values of the anisotropy  parameter $a$  are shown in
Figure~\ref{fig:anis}   (upper panels).   The  lower  panels  show the
corresponding   distributions of     apo-to-pericenter   ratios.  Each
distribution  is computed from a  Monte  Carlo simulation with  $10^6$
orbits. The thin vertical lines in Figure~\ref{fig:anis} show the 20th
(dotted lines), 50th (solid lines), and 80th (dashed lines) percentile
points of   the distributions.   The    average eccentricity  for  the
isotropic case ($a=0$)  computed from the  Monte Carlo  simulations is
$\overline{e}  \simeq 0.55$,  in excellent  agreement  with the  value
determined in Section~\ref{sec:analyt}. About 15 percent of the orbits
in  the isotropic singular isothermal sphere   have apo- to pericenter
ratios larger than $10$, whereas only $\sim 20$  percent of the orbits
have $r_{+}/r_{-}   <  2$.  Note however  that   these  numbers depend
strongly on the velocity anisotropy.

\subsection{Tracer populations}
\label{sec:tracers}

In the previous two  sections, we concentrated on the  self-consistent
case  of a  singular isothermal   halo  and the corresponding  density
distribution  that   follows   from  the  Poisson   equation.   Tracer
populations, however, do   not necessarily follow the  self-consistent
density  distribution.  Consider  a  tracer  population in a  singular
isothermal sphere potential with a density distribution given by
\begin{equation}
\label{tracerho}
\rho_{\rm trace}(r) = \rho_0 \left({r \over r_0}\right)^{-\alpha},
\end{equation}
(the self-consistent case   corresponds  to $\alpha =  2.0$).    If we
consider  the  same quasi-separable  DF   as in Section~\ref{sec:sing}
(i.e., equation~[\ref{quasidf}]), the  energy  dependence  of  the  DF
becomes
\begin{equation}
\label{tracedf}
g(E) = {\sqrt{\rm e} \, \rho_0 \, r_0^{\alpha} \over 4 \, \pi \, V_c^3
  \, u_{H,\alpha}} \; {\rm exp}\left[- {\alpha E \over V_c^2}\right],
\end{equation}
with
\begin{equation}
\label{uhalpha}
u_{H,\alpha} = \int\limits_{0}^{\infty} du \; {\rm e}^{(1-\alpha) u}
\int\limits_{0}^{\eta_{\rm max}} h_a(\eta) \; {\eta \; d\eta \over
\sqrt{\eta_{\rm max}^2 - \eta^2}}. 
\end{equation}

The   average     eccentricity can    be    computed   by substituting
equations~(\ref{tracedf}) and~(\ref{uhalpha}) in~(\ref{avs2}), thereby
using the expression of  $\Phi(r)$ given by  equation~(\ref{potdens}). 
In Figure~\ref{fig:trace}    we  plot the   average  eccentricity thus
computed   as a function  of the  power-law slope $\alpha$.  As can be
seen, the orbital eccentricities depend   only mildly on the slope  of
the density distribution.  Note that for $\alpha > 3$, the mass within
any radius $r>0$ is  infinite, whereas in  the case $\alpha < 3$,  the
mass outside any such radius is  infinite.  These differing properties
in  the  two  regimes  are apparent  in  the  behavior  of  the median
eccentricity seen   in Figure~\ref{fig:trace},  with the  intermediate
case of  $\alpha  = 3$  yielding  a minimum.    Due  to the infinities
involved the   cases examined may  not  accurately represent true dark
halos.  After giving an example of how such infinities cause problems,
the  subsequent sections we examine   other, more realistic, spherical
potentials with finite masses.

\placefigure{fig:trace}

\subsection{Stellar hydrodynamics and the virial theorem in the 
isothermal potential}
\label{sec:virial}

The  difference between infinite   and finite samples is evident  when
comparing the equation of stellar hydrodynamics to the virial theorem.
For  a sphere with  an isotropic DF,  the former takes the simple form
for the one dimensional velocity dispersion $\sigma$ :
\begin{equation}
\label{stellhydro}
{d~ \over dr} \left( \rho \sigma^2 \right) = -\rho { d\Phi \over dr}
\end{equation}
For  the   tracer  population  of   equation~(\ref{tracerho}) and   an
isothermal potential, this simplifies to:
\begin{equation}
\label{hydroans}
\sigma^2 =  -V_c^2  \left(  {d\,{\rm  ln}  \rho  \over  d\,  {\rm  ln}
    r}\right)^{-1} = {V_c^2 \over \alpha}
\end{equation}

In contrast, the  virial theorem states  that twice the kinetic energy
is equal to the virial. Adopting particle masses of unity yields:
\begin{equation}
\label{virialthm}
\sum v^2 = \sum \vec F \cdot \vec r = \sum {V_c^2 \over r} r 
\end{equation}
Since, the expectation value of $v^2$ is $3 \sigma^2$, this reduces
to:
\begin{equation}
\label{virialans}
\sigma^2 = {V_c^2 \over 3} 
\end{equation}
This  difference owes  to  the  assumption of finite   versus infinite
tracers.  The answers match only for $\alpha = 3$ where the divergence
in   mass is   only  logarithmic at  both   $r  \rightarrow 0$ and  $r
\rightarrow \infty $.    Even the ``self-consistent'' case  of $\alpha
=2$ has  a problem that is pointed  out in problem  [4-9] of Binney \&
Tremaine  (1987).  The kinetic  energy  per particle  is  $V_c^2$ in a
model with  only  circular orbits   and $ 3   V_c^2  /2$ if  they  are
isotropic.   Yet, they have the  same  potential and must satisfy  the
virial theorem.   In   Section~\ref{sec:omegacen}, we return  to  this
issue  as  we   examine  a case    where  the equations   of   stellar
hydrodynamics   have been   used   on astrophysical   objects  where a
subsample of a finite number of global tracers  has been observed.  We
now turn to the calculation  of the distribution of eccentricities for
finite sets of tracers.

\subsection{Truncated isothermal sphere with core}
\label{sec:trunc}

Given  the infinite  mass  and  central  singularity  of the  singular
isothermal   sphere,  dark  halos  are   often  modeled as  truncated,
non-singular, isothermal spheres with a core:
\begin{equation}
\label{trunciso}
\rho(r) = {M \over 2 \, \pi^{3/2} \, \gamma \, r_t} \;
{\exp(-r^2/r_t^2) \over r^2 + r_c^2}.
\end{equation}
Here  $M$, $r_t$, and $r_c$ are  the  mass, truncation radius and core
radius, respectively, and $\gamma$  is a normalization  constant given
by
\begin{equation}
\label{alpha}
\gamma = 1 - \sqrt{\pi} \; \left( {r_c \over r_t} \right) 
\exp(r_c^2 / r_t^2) \left[ 1 - {\rm erf}(r_c/r_t) \right].
\end{equation}
Since for  this density distribution the DF  is not known analytically
(i.e., this  requires the knowledge of  $\rho(\Phi)$ in order to solve
the  Eddington  equation), and since   this density distribution is no
longer scale-free,  we have to  use  a different approach  in order to
determine  the distribution of  orbital eccentricities.  We employ the
method described  by   Hernquist (1993).  We  randomly  draw positions
according to the density distribution.  The radial velocity dispersion
is computed from the second-order  Jeans equations, assuming isotropy,
i.e.  assuming $f = f(E)$:
\begin{equation}
\label{radv}
\overline{v_r^2}(r) = {1 \over \rho(r)} \int_r^{\infty} \rho(r) {d\Phi
  \over dr} dr.
\end{equation}
We compute local velocities $v$ by drawing  randomly a unit vector and
then a  magnitude  from a Gaussian   whose second  moment is equal  to
$\overline{v_r^2}$, truncated at the local escape speed.  Once the six
phase-space coordinates are known, the energy and angular momentum are
calculated, providing the apo- and pericenter  of the orbit by solving
equation~(\ref{rooteq}).   The    resulting  distributions  of orbital
eccentricities are not rigorous, since the velocity field has not been
obtained from  a  stationary DF,  but  rather  only used  the   second
moments.  However,  as demonstrated   by  Hernquist (1993),   $N$-body
simulations  run   from    these initial conditions  are     nearly in
equilibrium  (see also  Section~\ref{sec:simul}), which  suggests that
the eccentricity distributions derived are  sufficiently close to  the
actual equilibrium distributions.

\placefigure{fig:trunc}

In Figure~\ref{fig:trunc},  we  plot the   20th (dotted lines),   50th
(solid lines), and    80th (dashed lines)  percentile   points  of the
distributions of eccentricity   (left  panel) and apo-  to  pericenter
ratios (right panel), as functions of  $r_t/r_c$.  For $r_c = r_t$ the
distribution  of eccentricities is   almost symmetric, with the median
equal to $0.50$.   When $r_t/r_c$ increases,  the distribution becomes
more and   more   skewed  towards   high    eccentricity orbits;   the
distribution closely approaches that of the singular isothermal sphere
in the limit $r_t/r_c \rightarrow \infty$.

\placefigure{fig:jafhern}

\subsection{Steeper halo profiles}
\label{sec:nfw}

To examine the dependence of the orbital  eccentricities on the actual
density   distribution  of the  halos   we  determine the eccentricity
distributions of two well-known  models   with steeper outer   density
profiles ($\rho \propto r^{-4}$):
\begin{equation}
\label{rhojaf}
\rho_J(r) = {M \over 4 \pi} \; {a \over r^2 (r + a)^2},
\end{equation}
and
\begin{equation}
\label{rhohern}
\rho_H(r) = {M \over 2 \pi} \; {a \over r (r + a)^3},
\end{equation}
where $M$   is  the total mass.    These profiles  differ only  in the
steepness  of the  central cusp: the  former  one, known as the  Jaffe
(1983) profile, has a $r^{-2}$-cusp, whereas the  latter, known as the
Hernquist (1990) profile, has a  shallower $r^{-1}$-cusp.  We use  the
technique   described   in   Section~\ref{sec:trunc}   to compute  the
distributions of orbital eccentricities for isotropic DFs $f(E)$.

The results  are shown in Figure~\ref{fig:jafhern},  where we plot the
normalized eccentricity   distributions  for the Jaffe   and Hernquist
spheres. For comparison the results for the singular isothermal sphere
with isotropic DF are reproduced as well.  The three distributions are
remarkably similar (the differences are  best appreciated by comparing
the thin lines  indicating the percentile  points).  The distributions
are progressively skewed  toward higher eccentricities in the sequence
isothermal, $\rho_H(r)$, $\rho_J(r)$, but only moderately so.

Navarro,  Frenk \&  White (1995,  1996,  1997)  have used cosmological
simulations to  argue  that the  outer density profiles  of dark halos
decline as  $r^{-3}$.  Their profiles  are likely created by a variety
of numerical artifacts   (Moore  \etal 1998).  However,  our   results
suggest that such debates  will  not significantly alter the  expected
distributions  of eccentricities;   velocity  anisotropy is   far more
important than the details of the density profile.

\section{Orbital decay of eccentric orbits by dynamical friction}
\label{sec:friction}

The orbits of substructures  within  halos change owing to   dynamical
friction.  Chandrasekhar (1943)   derived the local deceleration of  a
body with mass $M$ moving  through an infinite and homogeneous  medium
of  particles with mass $m$.  The  deceleration is proportional to the
mass $M$, such that more massive sub-halos sink more rapidly.  As long
as $M \gg m$  the frictional drag is  proportional to the mass density
of  the medium, but  independent of the  mass $m$ of the constituents. 
For an isotropic, singular isothermal sphere the deceleration is
\begin{equation}
\label{dvdt}
{d \vec  v \over d   t} = -{G  M_s \over  r^2}  \; {\rm ln}\Lambda  \;
\left({v    \over V_c}\right)^{-2}  \left\{    {\rm  erf}\left({v\over
      V_c}\right) -  {2  \over \sqrt{\pi}}  \left({v\over  V_c}\right)
  {\rm exp}\left[  -\left({v\over    V_c}\right)^2   \right]  \right\}
\vec{e}_v
\end{equation}
with $M_s$  and  $v$  the  mass  and  velocity of  the  object   being
decelerated, $r$ the distance  of that object from  the center  of the
halo, ${\rm ln} \Lambda$ the   Coulomb logarithm, and $\vec{e}_v$  the
unit velocity vector (Tremaine 1976; White 1976a).

As mentioned earlier, remarkably little attention has been paid to the
effects of  dynamical friction on  eccentric orbits---our goal for the
rest  of this section.  Our  main objective is  to use both analytical
and numerical tools to investigate the  change of orbital eccentricity
with  time, and the dependence of  the dynamical friction timescale on
the (intrinsic) eccentricity.   Unlike most previous studies, we  will
not focus on testing Chandrasekhar's dynamical friction formula, or on
studying the exact cause of the deceleration  (i.e., local or global),
as this has been the topic of discussion in many previous papers.

\subsection{The time-dependence of orbital eccentricity}

We investigate  the rate at which orbital  eccentricity changes due to
dynamical  friction. For simplicity,    we focus on the   evolution of
orbital eccentricity in the singular isothermal sphere, for which
\begin{equation}
\label{dedt}
{d e \over d t} = {d \eta \over d t} \; {d e \over d \eta},
\end{equation}
with $\eta = L/L_c(E)$ (see Section~\ref{sec:sing}).  Using
equation~(\ref{circrad}), we find
\begin{equation}
\label{detadt}
{d \eta \over d t} = \eta \left\{ {1 \over L} \; {d L\over d t} - {1
    \over V_c^2} \; {d E \over d t}\right\}.
\end{equation}
Because of  dynamical friction the energy and  angular momentum are no
longer conserved, and
\begin{equation}
\label{denergydt}
{d E \over d t} =  v \; {d v \over d t}
\end{equation}
and
\begin{equation}
\label{dangmomdt}
{d L \over d t} =  r \; {d v_{\bot} \over d t}.
\end{equation}
Since the frictional force acts in the direction opposite of the
velocity
\begin{equation}
\label{dvperpdt}
{d v_{\bot} \over d t} = {v_{\bot} \over v} \; {d v \over d t}.
\end{equation}
and, upon combining equations~(\ref{dedt}) - (\ref{dvperpdt}), we find
\begin{equation}
\label{deccdt}
{d e \over d t} = {\eta \over v} \; {d e \over d \eta} \;
 \left[ 1 - \left({v \over V_c}\right)^2 \right] \; {d v \over d t}. 
\end{equation}
Here  $d    v /  d  t$   is   the frictional  deceleration   given  by
equation~(\ref{dvdt}). Since both $d v / d t < 0$ and $d e  / d \eta <
0$  (see Figure~\ref{fig:eps}), we can immediately  derive the sign of
$d e / d t$ at apo- and pericenter.  At apocenter $v < V_c$, such that
$d e / d t > 0$, whereas at pericenter $v > V_c$ and thus $d e / d t <
0$.  This explains the  circularization found for `grazing encounters'
(Bontekoe \& van Albada 1987), as dynamical friction happens only near
pericenter.   Equation~(\ref{dvdt})  shows  that $d  v  /  d t \propto
r^{-2}$, and the  change in eccentricity  is thus larger at pericenter
than at apocenter.  However, the time spent near pericenter is shorter
than near apocenter, so that the evolution of the eccentricity can not
be  determined by inspection.  In  the next sections, we use numerical
simulations.
  
\subsection{$N$-body simulations}
\label{sec:simul}

We perform a set of fully self-consistent  $N$-body simulations with a
large number of particles in order to examine the effects of dynamical
friction on eccentric orbits in a massive halo. The halo is modeled by
a truncated  isothermal sphere (equation~[\ref{trunciso}]), with total
mass of unity, a core-radius $r_c = 1$, and a truncation radius $r_t =
50$. Scaled  to  the    Milky Way, we    adopt  a  unit  of mass    of
$10^{12}\Msun$ and   a   unit of    length of   $4$  kpc.   With   the
gravitational constant set  to unity, the  units of  velocity and time
are $1037 \kms$ and $3.8$ Myr, respectively.
 
The initial  velocities of the halo particles  are set up as described
in  Section~\ref{sec:trunc}, following  the   procedure  of  Hernquist
(1993). Because of this particular method  the halo is not necessarily
in equilibrium, nor  is  it expected to  be  virialized.  In  order to
remove effects   of the  halo's virialization  on   the decay   of the
orbiting substructure, we first simulate  the halo in isolation for 10
Gyrs. At the end,  the halo has nicely  settled in virial equilibrium. 
Figure~\ref{fig:densprof}  shows the  initial density profile compared
to that  after 10 Gyr.  As can be seen, and  as already pointed out by
Hernquist (1993), the  density profile  has not changed  significantly
after 10 Gyr.

\placefigure{fig:densprof}

We are  interested  in the  effects of dynamical  friction on galactic
objects that range from $\sim 10^6 \Msun$ (globular clusters) to $\sim
10^{10} \Msun$ (a  massive satellite).  To simulate dynamical friction
on an   object of mass $M_s$  orbiting  in a halo   of  mass $M_h$, we
require $M_s \gg  m$.  In our simulations, we  insist that  $N \gta 10
M_h/M_s$.  A simulation of the orbital decay  of a globular cluster in
a galactic halo of $10^{12}\Msun$ thus requires $N \gta 10^7$, clearly
too large a  number  for practical  purposes.  We run  our simulations
with $N=50,000$ particles, roughly an order of magnitude increase over
most  previous    work.    The highest   resolution,   self-consistent
simulation  aimed at investigating dynamical friction  has so far been
performed by Hernquist   \&   Weinberg (1989) using  $N=20,000$    and
focusing only on   circular orbits.  We discuss  the  influence of the
number of halo particles in Section~\ref{sec:npart}.  With $N=50,000$,
$M_h  = 10^{12}\Msun$,  and the requirement  $N \gta  10 M_h/M_s$, our
minimum satellite   mass is  $2\times 10^8 \Msun$.    A  list  of  the
parameters  for   each    of  our    simulations   is  presented    in
Table~\ref{tab:param}.  Models~1, 2,  and 3 have an initial  apocenter
of   160 kpc,  and   start on  an   orbit  with  $e=0.8$.  The initial
pericenter  for these orbits is located  at $17.9$ kpc from the center
of the  halo, well outside the  halo's core radius   ($r_c = 4$  kpc). 
Satellites on  orbits of lower  eccentricity (Models~4, 5, and  6) are
started from smaller apocentric radii,  such that the initial specific
energy of all satellites in all models is equal.

The satellite is initially positioned at $(x,y,z) = (r_{+},0,0)$, with
$v_x  = v_z =  0$ and $v_y$ chosen such  as to obtain an initial orbit
with the desired eccentricity: we randomly draw velocities $0 \leq v_y
\lte v_{\rm escape}$ and   determine the orbital eccentricity   of the
satellite as  described   in  Section~\ref{sec:sing}.  We repeat  this
procedure until we  find  an eccentricity within  one  percent of  the
desired value. At  the  start of each   simulation, the satellite   is
introduced instantly in the virialized halo potential. 

The simulations use PKDGRAV (Dikaiakos \& Stadel 1996; Stadel \& Quinn
1998), a   stable  and  well-tested,  both   spatially and  temporally
adaptive tree  code, optimized for massively  parallel processors.  It
uses  an open ended variable  timestep  criteria based upon the  local
acceleration (Quinn  \etal 1997).  Forces are  computed using terms up
to  hexadecapole order and  a tolerance parameter of  $\theta  = 0.8$. 
The code uses  spline kernel softening,  for  which the  forces become
completely Newtonian at 2  softening  lengths (see Hernquist  \&  Katz
1989  for details).   In  terms  of where  the   force  is 50  percent
Newtonian,  the  equivalent Plummer softening  length  would be $0.67$
times the  spline softening length.  The softening  length of the halo
particles  is set  to $\epsilon   =  0.05$, or   $200$ pc.    Particle
trajectories are computed using a   standard second order   symplectic
leap-frog  integrator,  with  a   maximum time-step  $\Delta  t  =  1$
(corresponding to $3.77$ Myr in  our adopted physical units).  Because
of the multi-time stepping, some particles are integrated with smaller
timesteps.  For a   typical simulation, approximately  $20$ percent of
the particles  are  advanced with $\Delta  t =  0.5$, and  $\sim 0.05$
percent with $\Delta t = 0.25$.

The satellite  is  modeled as a single  particle  with mass $M_s$  and
softening length  $\epsilon_s$.  Beyond $2  \epsilon_s$ the  satellite
potential  falls as   $r^{-1}$,   so  this  radius approximates    the
satellite's tidal radius, which  is mainly determined by conditions at
pericenter (King 1962).   In  principle.  we could fix  the  softening
using an appropriate tidal radius.   However, the pericentric distance
evolves and  we would   then have to   include  time evolution  of the
softening and some mass  loss.  We opt for a  simpler approach and fix
the mean density of each satellite:
\begin{equation}
\label{softsat}
\epsilon_s = 2.73 \, {\rm  kpc} \; \biggl[  {M_s \over 10^{10}  \Msun}
\biggr]^{1/3},
\end{equation}
The scaling is set so that a satellite with mass  similar to the Large
Magellanic Cloud (hereafter LMC)  has a softening length comparable to
the LMC's effective  radius  (de Vaucouleurs  \& Freeman 1972).   This
choice  is  somewhat  arbitrary,  and  we  discuss its  influence   on
dynamical friction timescales in Section~\ref{sec:ressize}.

All simulations  are  run  for  $15$ Gyr  on   $2$ or  $3$ DEC   Alpha
processors,  each requiring about 48   hours of wallclock time. Energy
conservation was typically of the order of  one percent over the total
length of the simulation.

\placefigure{fig:orbits}

\subsection{Results}
\label{sec:resfric}

We determine  the eccentricity in two ways.  We compute the  center of
mass of the halo particles and use the  galactocentric distance of the
satellite  ($r_s$),  and its energy   and angular   momentum to  solve
equation~(\ref{rooteq}) for the orbital  eccentricity (shown as  solid
lines      in  Figures~\ref{fig:resa}    and~\ref{fig:resb}).     This
eccentricity is only accurate  if the potential  of  the halo has  not
changed significantly owing to the  satellite's decay.  Hence, we also
determine  the radial  turning points  of  the  orbit and compute  the
approximate eccentricity which we assign to  a time midway between the
turning   points  (shown  as open  circles   in Figures~\ref{fig:resa}
and~\ref{fig:resb}).

\placefigure{fig:resa}

In Figure~\ref{fig:orbits} we plot the  trajectories of the satellites
for models 1 to   6.  Both the   $x$--$y$ (upper panels)  and $x$--$z$
projections (smaller lower panels) are  shown.  The three trajectories
plotted on the top vary in satellite mass, whereas those at the bottom
vary      in    their    initial     orbital    eccentricity      (see
Table~\ref{tab:param}).    The time-dependences  of     galactocentric
radius, eccentricity, energy, and angular  momentum for various models
are  shown in Figures~\ref{fig:resa} and~\ref{fig:resb}.  Energies are
scaled by the central potential $\Phi_0$, and angular momenta by their
value at $t=0$.  Eccentricities are  only plotted up to $t_{0.8}$ (see
below),  after which  the  satellite has virtually  reached the halo's
center.     Models~1   and~2  reveal   an   almost  constant   orbital
eccentricity.  In Models~3 to~9, in which the  satellite mass is equal
to $2 \times 10^{10}    \Msun$, the eccentricity reveals  a  saw-tooth
behavior, such  that  eccentricities  decrease near   pericenter,  and
increase near    apocenter.     This is  in    perfect agreement  with
equation~(\ref{deccdt}).  It is remarkable that the net effect of $d e
/ d t$ is nearly zero: the eccentricity does not change significantly.
The  alternative definition of eccentricity  based on observed turning
points (open circles) shows similar results.  Small deviations are due
to a  change of the halo potential  induced by the decaying satellite,
and the  heuristic   assignment of a  time   to the  value  found from
monitoring the radial turning points.   The change in energy with time
reveals a step-wise behavior, indicating that the pericentric passages
dominate the  satellite's energy  loss.  Note that   the energy  of the
satellites in Models~3 to 9 does not become equal to $\Phi_0$ once the
satellite reaches the center of the potential  well.  This owes to the
deposition of  energy into the halo particles  by the  satellite.  The
details of this  process will be examined in  a future paper (van  den
Bosch \etal in preparation).

\placefigure{fig:resb}

Because of the elongation  of the orbits, the galactocentric  distance
is not a meaningful parameter to use to  characterize the decay times. 
Instead  we use both  the energy and  the angular momentum.  While the
energy   is  well defined,   it      changes  almost  stepwise    (see
Figures~\ref{fig:resa} and~\ref{fig:resb}).  The    angular   momentum
depends on the  precise position  of  the halo's  center which may  be
poorly determined when the satellite  induces an $m=1$ mode. We define
the   following    characteristic times:   $t_{0.4}$,  $t_{0.6}$,  and
$t_{0.8}$ defined as  the   time when the satellite's   energy reaches
$40$, $60$, and $80$ percent  of $\Phi_0$ respectively, and $t_{1/4}$,
$t_{1/2}$, and $t_{3/4}$ when  the  angular momentum  is reduced to  a
quarter,  a half  and  three-quarters  of  its initial  value.   These
timescales  are listed  in  Table~\ref{tab:timescale}.  Because of the
instantaneous introduction   of the satellite  in the  virialized halo
potential the absolute values of these timescales  may be off by a few
percent.   However, we are mainly  interested  in the variation of the
dynamical friction time  as function of  the orbital eccentricity.  We
belief that the  instantaneous introduction of  the satellite will not
have a  significant influence on  this behavior,  as  this is  only  a
second order effect.

\subsubsection{Influence of satellite mass}
\label{sec:resmass}

Models~1, 2, and 3  start on the same  initial orbit, but vary in both
the mass and size of the satellite.  Satellite masses correspond to $2
\times  10^8  \Msun$, $2  \times 10^9  \Msun$,  and  $2 \times 10^{10}
\Msun$  for Models~1,  2,  and~3   respectively.   The sizes of    the
satellites,  e.g.,   their   softening      lengths, are    set     by
equation~(\ref{softsat}), such that all  satellites have the same mean
density.  The  mass  and size of  the satellite  in model 3 correspond
closely  to that  of  the LMC.   It   is clear  from   a comparison of
Models~1,2, and 3   that dynamical friction by  the  galactic  halo is
negligible   for satellites with  masses  $\lta  10^{9} \Msun$.  Thus,
globular clusters  and the dwarf spheroidals in  the galactic halo are
not expected to have  undergone  significant changes in their  orbital
properties  induced by  dynamical friction by  the  halo; the only two
structures  in the galactic  halo   that have experienced  significant
amounts of  dynamical friction by  the halo  are the LMC  and the SMC,
with masses of  $\sim 2 \times 10^{10}  \Msun$ and $\sim 2 \times 10^9
\Msun$ respectively   (Schommer \etal  1992).     Note that  we   have
neglected the  action of  the disk and   bulge, which apply   a strong
torque  on objects passing nearby.    This can  result in an  enhanced
decay of   the orbit,  not   taken  into  account in   the simulations
presented here.

\subsubsection{Influence of orbital eccentricity}
\label{sec:resecc}

\placefigure{fig:halftime}

Models~3, 4, 5, and 6 differ only in the eccentricity of their initial
orbit; satellites  on less eccentric  orbits are started  from smaller
apocentric radii, such  that the initial  energy is  the same in  each
case.   The    characteristic    friction timescales    (defined    in
Section~\ref{sec:resfric}  and listed in Table~\ref{tab:timescale}) as
a    function     of      initial  eccentricity     are      shown  in
Figure~\ref{fig:halftime}.  The  decay  time decreases with increasing
eccentricity,  the exact rate  of which depends  on the characteristic
time employed.  The timescales for  the most eccentric orbit (Model~3)
are   a factor $1.5$ to  $2$   shorter  than for   the circular  orbit
(Model~6).  This exemplifies the importance  of  a proper treatment of
orbital eccentricities,  since  decay times  based on  circular orbits
alone overestimate the timescales   of dynamical friction for  typical
orbits in a dark halo by up to a factor two.

Using Chandrasekhar's dynamical friction formula, and integrating over
the orbits, Lacey \& Cole (1993) found that, for a singular isothermal
sphere, the  dynamical friction time  is proportional to $\eta^{0.78}$
(with      $\eta$     the    orbital       circularity    defined   in
Section~\ref{sec:analyt}).        The      dotted        lines      in
Figure~\ref{fig:halftime} correspond to this dependence, normalized to
$t_{0.6}$ and $t_{1/2}$ for  the orbit with an  intrinsic eccentricity
of $0.6$.   As  can be seen, our   results seem to suggest  a somewhat
weaker  dependence of    the  dynamical  friction    time on   orbital
eccentricity.  Combining  all the different characteristic decay times
listed in  Table~\ref{tab:timescale}, we obtain   the best fit  to our
$N$-body  results for  $t   \propto   \eta^{0.53 \pm  0.01}$.     This
difference with respect to the results of Lacey \& Cole is most likely
due to the fact that their  computations ignore the global response of
the halo to the decaying satellite.

\subsubsection{Influence of satellite size}
\label{sec:ressize}

\placefigure{fig:halfeps}

Since the adopted softening  length of a satellite, ${\epsilon_s}$, is
somewhat   arbitrary (see  Section~\ref{sec:simul}),   we  examine the
effect of this choice by running a set of models (3, 7, 8, and 9) that
differ only in this  quantity. The characteristic friction  timescales
as a function of ${\epsilon_s}$ are shown in Figure~\ref{fig:halfeps}.
In Chandrasekhar's formalism, the dynamical friction timescale depends
on the inverse of  the Coulomb logarithm ${\rm  ln} \Lambda$.  This is
often  approximated by the logarithm of  the  ratio of the maximum and
minimum impact parameters  for the satellite.   Taking $b_{\rm  max} =
r_t  = 200$ kpc and  $b_{\rm  min} = \epsilon_s$,  the expected ratios
between the timescales  of Models~3, 7, 8,  and  9 can  be determined. 
The dotted  lines in  Figure~\ref{fig:halfeps} shows these predictions
scaled to  the   characteristic  times of Model~3.   It   is apparent,
however, that these simulations are not accurately represented by this
simple approximation.   This is not surprising  as the choices for the
minimum and maximum impact  parameters are somewhat arbitrary. Because
of the radial density profile of the  halo, $b_{\rm max}$ is likely to
depend on radius.   Furthermore, approximating  $b_{\rm  min}$ by  the
satellite's softening length is  only appropriate for large  values of
$\epsilon_s$;  in the limit where the   satellite becomes a point mass
$b_{\rm min}$  should be taken as $G   M_s/\langle v^2 \rangle$, where
$\langle v^2 \rangle^{1/2}$  is the rms  velocity of the background, a
quantity which again   depends on the  galactocentric  distance of the
satellite (White 1976b).   The predictions in Figure~\ref{fig:halfeps}
are  thus as good  as we might have  hoped for. The important point we
wish to make here is that deviations  from the softening lengths given
by     equation~(\ref{softsat}),  within reasonable    bounds,  do not
significantly modify the  characteristic dynamical friction timescales
presented here.

\subsubsection{Influence of number of halo particles}
\label{sec:npart}

\placefigure{fig:comp}

In order  to address the accuracy  of our simulations we have repeated
Model~4 with both 20,000   and 100,000 halo particles.   We henceforth
refer to    these simulations as  Models~10 and~11   respectively (see
Table~\ref{tab:param}).  Both these   models are first evolved for  10
Gyrs   without satellite  to   let   the   halo virialize and    reach
equilibrium.   As   for the halo   with  50,000 particles, the density
distribution after 10 Gyrs has not changed significantly.

A  comparison  of  the results  of Models~4,   10  and~11 is  shown in
Figure~\ref{fig:comp}.  The   solid lines correspond to   Model~4, the
dotted lines   to Model~10 ($N =   20,000$), and  the dashed  lines to
model~11 ($N = 100,000$). Clearly our  results are robust, with only a
negligible  dependence on  the number of    halo particles.  The  main
differences   occur at later  times, when  the satellite has virtually
reached the  center  of the halo.  In  Models~4 and  11 the  satellite
creates a core in the halo, which causes $E/\Phi_0 < 1$ in the center.
Model~10  has insufficient  particles  to  resolve this  effect, which
explains   the  differences    in both the     satellite's  energy and
eccentricity at    later times with  respect   to models~4 and~11.  In
Table~\ref{tab:timescale}  we   list  the  different    characteristic
timescales for Model~10 and~11.  They are similar  to those of Model~4
to an  accuracy of $\sim 4$ percent  for Model~10 and $\sim 1$ percent
for Model~11. Finally we note that our main  conclusion, that there is
no net amount of circularization, holds for  different numbers of halo
particles.

\section{Applications}
\label{sec:applic}

\placefigure{fig:prob}

\subsection{Orbits of Globular clusters}
\label{sec:globulars}

Odenkirchen \etal (1997,   hereafter  OBGT)  used Hipparcos data    to
determine the proper  motions of 15  globular clusters so that all six
of their phase-space coordinates are known. Their velocity dispersions
show a  slight radial   anisotropy with $(\sigma_r,   \sigma_{\theta},
\sigma_{\phi}) = (127 \pm 24,  116 \pm 23, 104  \pm 20)$ $\kms$.  OBGT
integrated the orbits using a model  for the galactic potential (Allen
\& Santillan 1991) that approaches an isothermal  at large radii. They
find a  median  eccentricity of  $e =  0.62$  and  conclude  that  the
globulars are {\it preferentially on orbits of high eccentricity}

We can  compare  their  results  with  the eccentricity  distributions
expected for  a power-law tracer population  in  a singular isothermal
halo.  We use the technique described in the Appendix to determine the
distribution  of orbital eccentricities  given  a logarithmic slope of
the density distribution $\alpha$ and an anisotropy parameter $a$.  We
compare the  cumulative distribution  to  OBGT's sample using  the K-S
test (e.g., Press \etal 1992). We determine the probabilities that the
OBGT sample  is drawn randomly from such  a  distribution using a $100
\times 100$ grid in $(\alpha,a)$-space with $\alpha  \in [2,6]$ and $a
\in [-2,2]$   and show  the contour  plot  of these   probabilities in
Figure~\ref{fig:prob}.          The  contours  corresponds   to    the
$10,20,...,80,90$ percent  probability levels,  whereby the latter  is
plotted as  a  thick contour.   Clearly, the    slope of the   density
distribution,  $\alpha$, is poorly  constrained  as the dependence  is
mild   (see Section~\ref{sec:tracers}).    However,    the    velocity
anisotropy is  well constrained, and  we find that  a small  amount of
radial anisotropy    is  required in order   to   explain the observed
eccentricities   of  the globular  clusters.    This  is in  excellent
agreement  with the velocity dispersions   obtained directly from  the
data.  The solid dot in Figure~\ref{fig:prob}  corresponds to the best
fitting model  with $\alpha  =  3.5$, the  value preferred   by Harris
(1976) and Zinn   (1985).    In Figure~\ref{fig:ksplot} we   plot  the
cumulative  distribution of  the eccentricities  from  the OBGT sample
(thin     lines) with  the cumulative    distribution   for the  model
represented by the dot in  Figure~\ref{fig:prob}.  The K-S test yields
a  probability of $94.3$  percent that  the  OBGT sample of  orbits is
drawn  randomly from the   probability distribution represented by the
thick line.

\placefigure{fig:ksplot}

We conclude that the  distribution of eccentricities  is just what one
expects from a  population   with a mild radial   velocity anisotropy;
there is no ``preference''  for high eccentricity orbits as  suggested
by  OBGT.  The  potential  used  by  OBGT  to integrate  their  orbits
deviates  significantly from a spherical    isothermal in the  center,
where the disk and bulge  dominate.  Since the  OBGT sample is limited
to  nearby globulars  within approximately   20  kpc of  the Sun,  the
deviations are likely significant. However, the good agreement between
the  distribution of eccentricities for  the OBGT sample, based on the
axisymmetric potential  of Allen \& Santillan  (1991), and a spherical
isothermal  potential suggests that the  differences between these two
potentials have only a  mild influence on  the distribution of orbital
eccentricities.

\subsection{Kinematics in $\omega\;$Centauri}
\label{sec:omegacen}

Norris \etal  (1997)  examined the dependence  of  kinematics on metal
abundance   in the  globular    cluster $\omega\;$Centauri  (hereafter
$\omega\;$Cen) and found  that the  characteristic velocity dispersion
of the most  calcium rich stars is $\sim   8 \kms$, while that of  the
calcium  poor stars is  $\sim 13   \kms$.   The metal  rich  stars are
located closer to the middle where the  velocity dispersion is largest
and the  authors  note  that  there is evidence  for  rotation  in the
metal-poor sample  (at $\sim   5 \kms$), but   not in  the metal  rich
sample.  They  use all these  facts  to conclude that  ``{\it The more
  metal-rich stars in $\omega\;$Cen  are kinematically cooler than the
  metal-poorer objects}''.  The metal-rich stars  in their sample live
in the  part of the cluster where   the inferred circular  velocity is
greatest.  Hence, they have an average value of $\vec  F \cdot \vec r$
that is greater than for the metal-poor stars.  By the virial theorem,
their mean kinetic energy must be higher (equation~[\ref{virialthm}]),
yet Norris \etal  find just   the opposite  with  the average  kinetic
energy of a metal-poor star being more than twice that of a metal-rich
star.

Since we are not   about to abandon the virial   theorem, we can  only
conclude that {\it if the measurements of the dispersions are correct,
  the kinetic energy per metal-rich star must be at  least that of the
  metal-poor stars implying a much greater kinetic energy in the plane
  perpendicular to the  line of sight  than observed along the line of
  sight}.  The straightforward way   for this to  occur is  a rotating
disk seen face-on.  The rotation  must be large enough that $v_{rot}^2
+ 3 \sigma^2$ is  at least as large as  seen for the metal-poor stars. 
This implies that  the metal-rich stars   have a rotation  velocity of
$\gta 18 \kms$.   The metal-rich component  of $\omega\;$Cen must be a
rotating  disk that  is more  concentrated than the  metal-poor stars. 
This is exactly the signature that Norris \etal would have ascribed to
self-enrichment of the cluster   (Morgan \& Lake 1989).   The rotation
signature would  be visible as a  proper motion of $0.064 ''$/century. 
Rotation of a smaller  magnitude has been  detected in M22 by Peterson
\& Cudworth (1994).  Note that a face-on  disk in $\omega\;$Cen, which
is  the  Galactic  globular  with  the largest  projected  flattening,
implies a triaxial potential.

Norris  \etal did not  realize  that their observations presented this
dynamical puzzle.  Instead,   they believed that the difference  could
result  from  the relative  radial profiles  of  the two components as
might  be   seen   in   the   equations  of     stellar  hydrodynamics
(equation~[\ref{hydroans}]);  i.e.,  the  density distribution  of the
metal rich component must  fall off more  rapidly with radius than for
the  metal poor component.  They  used  the observations to argue that
$\omega\;$Cen was  the product of  a merger of previous generations of
substructure.  However, we argue that such a merger would not have the
signatures that they  see.  The metal-poor  stars are the overwhelming
majority of the stars.   Conservation of linear momentum thus  implies
that the mean radius of the stars that were  in the small (metal-rich)
object will be greater than those that  were in the large (metal-poor)
one.  Conservation  of  angular momentum furthermore implies  that the
rotation velocity  of the  stars that were   in the small (metal-rich)
object will be greater than those  that were in the large (metal-poor)
one.  These signatures are exactly opposite of those claimed by Norris
\etal to be consistent with the merger model.

\subsection{Sinking satellites and the heating of galaxy disks}
\label{sec:satellite}

The sinking  and subsequent merging of  satellites  in a galactic halo
with an embedded thin disk has been studied  by numerous groups (e.g.,
Quinn \& Goodman  1986; T\'oth \& Ostriker  1992; Quinn,  Hernquist \&
Fullager 1993;  Walker, Mihos  \& Hernquist  1996; Huang   \& Carlberg
1997).  The timescale for merging clearly  depends on the eccentricity
of its orbit.  Once the satellite interacts with the disk, the sinking
accelerates.    Several   of   the   above   studies    assumed   that
circularization owing to dynamical  friction by the halo  is efficient
and examined satellites that started on circular orbits at the edge of
the  disk.  However, we  have shown that  circularization is largely a
myth; satellites will have  large eccentricities  when they reach  the
disk.  Quinn \& Goodman (1986) followed a satellite with a ``typical''
eccentricity in an isotropic singular isothermal sphere.  They derived
$e \simeq 0.47$ or $r_{+}/r_{-}   = 2.78$ for this  ``characteristic''
orbit based on some poorly founded arguments.  This ratio, however, is
significantly smaller than the  true median value of $r_{+}/r_{-} \sim
3.5$; approximately $63$ percent of the orbits have $e > 0.47$.  Huang
\& Carlberg (1997), in an attempt to be as  realistic as possible when
choosing the initial  orbital parameters, started their satellite well
outside the disk on an  eccentric orbit.  However, the eccentricity is
only 0.2, clearly too low to be considered a typical orbit.

The eccentricity  of the orbit  has two effects; more eccentric orbits
decay more rapidly in  the halo and they touch  the disk at an earlier
time.  The  sinking time  scales   for  the satellite caused   by  its
interaction with  the   disk are more   rapid when  the difference  in
velocities  between  the  satellite and the   disk stars  is  smaller;
satellites on prograde, circular orbits in the disk decay fastest (see
Quinn \& Goodman 1986 for  a detailed discussion).  Thus whereas  more
eccentric orbits reach the disk sooner, they are less sensitive to the
disk interaction because of their high velocities at pericenter.  Less
eccentric orbits,  on the other  hand, require a  longer time to reach
the disk, but once they do  their onward decay is  rapid (if the orbit
is prograde).  The exact dependence of the timescales and disk heating
on the orbital eccentricities awaits simulations (van den Bosch \etal,
in preparation).

We can compare our results to the few satellite orbits determined from
proper-motions.  Johnston (1998)   integrated the orbits  of  the LMC,
Sagittarius,  Sculptor, and Ursa Minor  in the galactic potential, and
found   apo- to pericenter  ratios  of $2.2$,  $3.1$, $2.4$, and $2.0$
respectively.   Using the K-S test,  we find that these eccentricities
are so small that there is only a  $8.7$ percent probability that this
sub-sample is drawn from the isotropic model of  an isothermal sphere. 
As  shown   in  previous sections    the   distribution is  relatively
insensitive to the profiles  of    the underlying potential and    the
density distribution of the  tracer population. This therefore implies
that   the     velocity       distribution   is     very      strongly
tangential\footnote{Even if we adopt the  maximum amount of tangential
  anisotropy allowed  by  our simple parameterization  of  $h_a(\eta)$
  (i.e., $a  \rightarrow \infty$) the K-S  probability does not exceed
  20 percent.}.   Since we expect  that collapsed  halos would produce
states with dispersions   that are preferentially radial,  we  suspect
that {\it the system of  galactic satellites has been strongly altered
  with satellites on more eccentric orbits having been destroyed owing
  to    their    faster  dynamical     friction   timescales}.      In
Section~\ref{sec:friction}, we  found that  more eccentric orbits have
smaller  dynamical friction timescales,  but the effect is only modest
(less than a factor two).  Furthermore, all  the satellites except the
Magellanic  clouds  have masses $\lta   10^9 \Msun$, and the dynamical
friction owing to the halo is almost negligible for these systems.  To
get the strong effect that is seen, we have to appeal to the Milky Way
disk to accelerate the decay and/or  tidally disrupt the satellites on
the more eccentric orbits. The problem with this solution, however, is
that most studies  of sinking satellites have  shown that they lead to
substantial thickening  of the  disk.   Further studies  are needed to
investigate whether a disk  can indeed yield the observed distribution
of orbital eccentricities for the (surviving) satellites without being
disrupted itself.  We   are currently using  high  resolution $N$-body
simulations  to investigate this is detail  (van  den Bosch \etal , in
preparation).  Finally, we  must  note that the sample  of  satellites
with proper  motions is small and  the  proper motions themselves have
large errors which   bias results  toward  large  transverse  motions. 
Hence, we close  this section with the  all too  common lament of  the
need  for more data with more  precision as well as better simulations
that include the disk.

\subsection{Tidal streams}
\label{sec:streams}

The tidal disruption of satellites orbiting in a galactic halo creates
tidal streams (see e.g., McGlynn 1990; Moore \& Davis 1994; Oh, Lin \&
Aarseth 1995; Piatek \& Pryor   1995; Johnston, Spergel \&   Hernquist
1995).  These streams are  generally long-lived features outlining the
satellite's orbit  (Johnston,  Hernquist \&  Bolte  1996).  Clearly, a
proper  understanding of the  orbital  properties of satellites is  of
great importance when studying tidal streams and estimating the effect
they might have on the statistics of micro-lensing (i.e., Zhao 1998).

An   interesting  result  regarding tidally   disrupted satellites was
reached by Kroupa (1997) and  Klessen \& Kroupa (1998). They simulated
the tidal disruption  of  a satellite without  dark matter  orbiting a
halo.  After tidal  disruption, a stream  sighted along  its orbit can
have  a spread   in velocities  that  mimics   a bound  object with  a
mass-to-light ratio  that can be  orders of  magnitude larger than the
actual, stellar mass-to-light   ratio.   They show that  one  can only
sight down such a  stream if the satellite  was on an eccentric orbit. 
The chance   of inferring  such  a   large mass-to-light  ratio   thus
increases with eccentricity.  Klessen \& Kroupa conclude that the high
inferred mass-to-light ratios   in observed dwarf spheroidal  galaxies
could occur  from tidal streams  (rather than from  a dark matter halo
surrounding the satellite)  {\it if the  orbital eccentricities exceed
  $\sim 0.5$}.  We   find that $\sim 60$\%   of the  orbits  obey this
criterion   even    without    any   radial     anisotropy      (e.g.,
Figure~\ref{fig:anis}).  We  can thus not  rule out satellites without
dark matter  based on the distribution  of orbits. However, Klessen \&
Kroupa (1998)  neglect  to  examine  the kinematics  in  detail.   The
velocity spread owes  to a systematic  gradient in velocity  along the
stream.  The apocryphal dwarf will  appear to rotate  rapidly if it is
not perfectly  aligned  with the line of  sight,  as is  clear  in the
simulations of the  Sagittarius   dwarf performed  by  Ibata \&  Lewis
(1998). 

\placefigure{fig:ghigna}

\subsection{Clusters of galaxies}
\label{sec:clusters}

Using very  high resolution cosmological  $N$-body simulations, Ghigna
\etal  (1998) were able to  resolve several  hundred subhalos within a
rich cluster of galaxies.   They examined, amongst others, the orbital
properties of these dark  matter halos within  the larger halo of  the
cluster, and found that the subhalos followed the same distribution of
orbits as the  dark matter  particles  (i.e., those  particles in  the
cluster that are not part of a subhalo).  Ghigna \etal report a median
apo-to-pericenter ratio of six. We have used the orbital parameters of
the  subhalos of the cluster analyzed  by Ghigna  \etal to examine the
orbital properties in some more detail. When we only consider subhalos
whose apocenters are   less  than the  virial radius   of the cluster,
$r_{200}$, we obtain a   sample of $98$  halos with  a median apo-  to
pericenter    ratio   of  $3.98$.   Using      the K-S   test   as  in
Section~\ref{sec:globulars} we obtain a  best fit to  the distribution
of orbital  eccentricities for an   anisotropy parameter of  $a \simeq
-0.04$:  the   virialized region   of the cluster    is very  close to
isotropic.    Throughout we  consider only  a  tracer population  with
$\alpha = 2$, since this is in reasonable  agreement with the observed
number density of  subhalos.  Furthermore, the  results presented here
are very insensitive to  the exact value  of $\alpha$, similar to what
we  found in Section~\ref{sec:globulars}.  In  Figure~\ref{fig:ghigna}
we  plot the cumulative  distribution of the orbital eccentricities of
the 98 halos with $r_{+} < r_{200}$ (thin line) together with the same
distribution for an isothermal sphere with our best fitting anisotropy
parameter $a = -0.04$.   The K-S test yields  a probability  of $61.3$
percent that the two data sets are drawn from the same distribution.

When analyzing all subhalos with  $r_{+} < 2  \, r_{200}$, we obtain a
sample of  311 halos with  a median apo-to-pericentric ratio of $4.64$
and a best fitting value for the anisotropy  parameter of $a = -0.27$. 
Clearly, the periphery of the cluster, which is not yet virialized, is
more radially anisotropic with more orbits on more eccentric orbits.

The $N$-body  simulations in Section~\ref{sec:simul} are easily scaled
to clusters of galaxies,  considering  the virialized regions  of both
galaxies and clusters. If we adopt  a cluster mass of $10^{15} \Msun$
and take the timescale to be the same as for the Milky Way simulation,
the     truncation radius      becomes   $r_t     =  2$   Mpc.      In
Section~\ref{sec:resmass} we   found  that only  objects with   masses
greater than $\sim  0.1$ percent of  the halo mass, $M_{\rm gal}  \gta
10^{12} \Msun$   in   the cluster,  experience  significant  dynamical
friction in  a   Hubble time.  So,    only the most   massive galaxies
experience any orbital decay.

Recently, Moore  \etal  (1996)  showed  that  high  speed   encounters
combined  with global  cluster tides---galaxy harassment---causes  the
morphological transformation of  disk systems into spheroids (see also
Moore, Lake  \& Katz 1998).  Moore \etal  limited themselves to mildly
eccentric orbits with $r_{+}/r_{-}  =  2$, but they quoted   correctly
that this was a low value compared to the typical eccentricity.  Their
choice was made in  order to be  conservative and to underestimate the
effect    of harassment,  as     the  effect  increases   with orbital
eccentricity.  They also felt that any larger  value would stretch the
reader's credulity as they could refer to no clear presentation of the
likely distribution of orbital eccentricities.  Our results imply that
the effects of harassment were underestimated in their study.

\subsection{Semi-analytical modeling of galaxy formation}
\label{sec:sam}

Over    the  past couple  of    years several   groups have  developed
semi-analytical models for galaxy formation  within the framework of a
hierarchical   clustering  scenario   of structure   formation  (e.g.,
Kauffmann, White \& Guiderdoni 1993; Cole \etal 1994; Heyl \etal 1995;
Baugh, Cole  \& Frenk 1996; Somerville  \& Primack 1998).  The general
method   of  these  models is  to   use the   extended Press-Schechter
formalism (Bower 1991;  Bond \etal 1991; Lacy  \& Cole 1993) to create
merging  histories  of  dark matter   halos.  Simplified yet  physical
treatments  are subsequently used to   describe  the evolution of  the
baryonic gas component in these halos.  Using  simple recipes for star
formation and feedback, coupled to  stellar population models, finally
allows predictions for    galaxies to be   made  in  an  observational
framework.

A crucial ingredient of these semi-analytical  models is the treatment
of mergers of  galaxies. When two dark  halos merge, the fate of their
baryonic cores, i.e.,  the galaxies, depends on  a number  of factors. 
First of all, dynamical friction causes the galaxies  to spiral to the
center of the combined dark halo, thus  enhancing the probability that
the baryonic cores collide.  Secondly, whether or not such a collision
results  in  a merger depends  on the  ratio  of the internal velocity
dispersion  of the galaxies involved  to the  encounter velocity.  Both
depend critically  on the     masses  involved and on   the    orbital
parameters.  The dependence on the orbital eccentricities is addressed
by Lacey \&   Cole (1993) who concluded  that   observations of merger
rates   and the  thinness of   galactic disks  seem  to argue  against
strongly elongated orbits.  This would be problematic  in the light of
the typical  distributions of orbital   eccentricities presented here. 
However, as we pointed  out in Section~\ref{sec:resecc}, Lacey \& Cole
have likely overestimated  the dependence of  dynamical friction times
on    orbital    eccentricity.     Furthermore,     as emphasized   in
Section~\ref{sec:satellite}  our current understanding of the damaging
effect of sinking satellites on thin disks is not well established and
may have been overestimated in the past (Huang \& Carlberg 1997).

\placefigure{fig:avertime}

In the  actual  semi-analytical modeling  the  merger time-scales  are
defined by simple scaling laws  that depend  on  the masses only,  but
that ignore the   orbital parameters.  The  eccentricity distributions
presented here, coupled with the $\eta^{0.53}$ dependence of dynamical
friction times, may  proof helpful  in improving  the accuracy of  the
merging  timescales  in  the   semi-analytical  treatments   of galaxy
formation. As    an  illustrative example,  we  calculate  the average
dynamical friction time
\begin{equation}
\label{frictime}
\langle t/t_0 \rangle = \int\limits_{0}^{1} d\eta \; \eta^{0.53} \,
p_a(\eta),
\end{equation}
where $t_0$ is the friction time  for a circular orbit and $p_a(\eta)$
is the normalized distribution  function of orbital circularities in a
singular  isothermal  sphere  with anisotropy   parameter  $a$.   This
average   time      is    plotted    as   function     of      $a$  in
Figure~\ref{fig:avertime} (solid line). For comparison we also plotted
the average   times obtained by  assuming  a  $\eta^{0.78}$ dependence
(dashed line).   The average dynamical friction   time is typically of
the order of $70$  to $80$ percent of that  of the circular orbit. The
stronger dependence  of   Lacey \&  Cole underestimates  the   typical
friction times by approximately $10$ percent.

\section{Conclusions \& Discussion}
\label{sec:conc}

This paper  has presented the  distributions of orbital eccentricities
in   a variety of   spherical  potentials.  In  a  singular isothermal
sphere,  the  median   eccentricity  of an   orbit  is   $e  =  0.56$,
corresponding to an  apo-  to pericenter ratio  of 3.55.    About $15$
percent of the orbits have $r_{+}/r_{-} >  10$, whereas only $\sim 20$
percent have  moderately   eccentric orbits with  $r_{+}/r_{-}  <  2$. 
These values depend strongly  on the velocity anisotropy  of the halo. 
Collapse is  likely  to create radially biased   velocity anisotropies
that  skew the distribution to   even higher eccentricities.  We  also
examined the  distributions  of orbital   eccentricities of  isotropic
tracer populations with a  power-law density distribution  ($\rho_{\rm
  tracer} \propto r^{-\alpha}$)  and found  only modest dependence  on
$\alpha$.    Due to the  unphysical nature  of the singular isothermal
sphere, we examined more realistic models and found that they differed
only slightly from the isothermal case.

We stress that these  eccentricity distributions apply only to systems
in equilibrium.   If a tracer  population has not yet fully virialized
in   the  halo's  potential,     its orbital   eccentricities can   be
significantly  different  from the    virialized  case.   Cosmological
simulations of  galaxy clusters and  satellite systems around galaxies
show prolonged  infall,   and recent mergers   can produce  correlated
motions.   Hence, care must  be taken in  applying our results to such
systems.

Objects with mass fractions  greater than 0.1\% experience significant
orbital decay owing to  dynamical friction.   We used high  resolution
$N$-body simulations with $50,000$   particles to examine  the sinking
and (lack of)   circularization of eccentric  orbits in   a truncated,
non-singular isothermal halo.  We derived, and numerically verified, a
formula that describes the change of eccentricity with time; dynamical
friction increases the eccentricity  of an orbit near  pericenter, but
decreases   it again  near   apocenter, such  that   no net amount  of
circularization occurs. The energy loss owing to dynamical friction is
dominated  by the deceleration  at  pericenter resulting in moderately
shorter  sinking  timescales for   more eccentric  orbits.  We find  a
dependence of the  form $t \propto  \eta^{0.53}$;  the average orbital
decay time for an isotropic, isothermal sphere is $\sim 75$ percent of
that  of  the  circular orbit.  This   dependence  is weaker  than the
predictions of Lacey \& Cole (1993), who found $t \propto \eta^{0.78}$
based on analytical integrations of Chandrasekhar's dynamical friction
formula.  Since this analytical  treatment ignores the global response
of the halo to the  decaying satellite we believe   our results to  be
more accurate.  This  relatively   weak dependence of decay   times on
eccentricity, together  with the absence  of any significant amount of
circularization, implies  that   dynamical friction does not  lead  to
strong changes in the   distribution of orbital  eccentricities.  When
scaling the simulations to represent the orbiting of satellites in the
galactic halo, we find  that the LMC and the  SMC are the only objects
in the outer Milky Way halo that  have experienced significant amounts
of energy and angular momentum loss by dynamical friction.

The distribution  of  orbital eccentricities is important  for several
physical   processes   including:  timescales  for   the   sinking and
destruction of galactic satellites, structure and evolution of streams
of tidal  debris, harassment   in   clusters of galaxies,  and    mass
estimates based  on the dynamics of the  system of  globular clusters. 
The results presented here may proof particularly useful for improving
the treatment  of galaxy mergers in  semi-analytical  models of galaxy
formation.   In    Section~\ref{sec:applic}   we    showed   that  the
distribution of orbital eccentricities of  a subsample of the galactic
globular cluster system is consistent with that of a slightly radially
anisotropic $r^{-3.5}$ tracer  population in an isothermal potential.  
A similar result was found  for the subhalos  orbiting a large cluster
of galaxies in a  high  resolution, cosmological $N$-body   simulation
presented by Ghigna \etal  (1998).  However, the Milky  Way satellites
are not consistent  with this  distribution, but  show a bias   toward
circularity that   may have been  caused by  dynamical friction and/or
tidal disruption by the galactic disk.  We expect that additional data
together with simulations that include the disk will lead to stringent
constraints on  the  formation and evolution   of  substructure in the
Milky Way.


\acknowledgments

We are  grateful to  Sebastiano  Ghigna  for  sending  us his data  in
electronic form.  We  are indebted to  Derek Richardson,  Jeff Gardner
and  Thomas   Quinn for  their  help  and  support with   the $N$-body
simulations, and to  the  anonymous referee  for his helpful  comments
that improved the  paper.  FvdB was supported  by a Hubble Fellowship,
\#HF-01102.11-97A, awarded by STScI.


\clearpage

\appendix

\section{Monte Carlo method to compute distributions of orbital
  eccentricities in a singular isothermal potential}
\label{sec:AppA}

To determine the distribution of orbital eccentricities in a spherical
potential,  one  must  sample  orbits  according  to  the distribution
function $f(E,L)$. In this appendix,  we describe a Monte Carlo method
that     samples      a   quasi-separable      distribution   function
(equation~[\ref{quasidf}]).  Once  the energy and angular  momentum of
an   orbit  are   known,    the eccentricity  is    easily  determined
(Section~\ref{sec:sing}).  In the following,   ${\cal R}$ is a  random
number  in  the interval   $[0,1]$,  ${\cal P}$ denotes  a probability
distribution, and $\widehat{f}$  indicates  that the function  $f$  is
normalized.

For a singular isothermal  density distribution with a quasi-separable
distribution function of  the form given  by equation~(\ref{quasidf}),
the density can be written as
\begin{equation}
\label{densiso}
\rho(r) = {4 \pi \over r} \int\limits_{\Phi(r)}^{\infty} dE \; g(E) \;
L_c(E) \int\limits_0^{\eta_{\rm max}} h(\eta)  \; {\eta \; d\eta \over
  \sqrt{\eta_{\rm max}^2 - \eta^2}}.
\end{equation}
The  joint probability distribution  of  $(E,\eta)$ at a fixed  radius
$r_0$ is therefore
\begin{equation}
\label{jointprob}
{\cal P}(E,\eta) =  {4 \pi \over r_0  \rho(r_0)} \; g(E) \;  L_c(E) \;
{\eta \; h(\eta) \over \sqrt{\eta_{\rm max}^2 - \eta^2}},
\end{equation}
with $E >  \Phi(r_0)$ and $\eta \in [0,\eta_{\rm  max}]$ (cf.  van der
Marel, Sigurdsson \& Hernquist 1997).

Because  of   the quasi-separable nature   of the  DF, the probability
function for the energies $E$ can be separated:
\begin{equation}
\label{prob_ener}
{\cal P}(E) = {4 \pi \over r_0 \rho(r_0)} g(E) L_c(E) = {r_0 \over u_H
  V_c^2} \; \exp\left[ -{E \over V_c^2} \right].
\end{equation}
The normalized, cumulative probability distribution of the energy is
\begin{equation}
\label{prob_ener_norm}
\widehat{\cal P}( < E)  = 1 -  \exp\left[ {\Phi(r_0) - E  \over V_c^2}
\right].
\end{equation}
Using the inversion of $\widehat{\cal P}( < E)$, energies are selected
by drawing a random number ${\cal R}$ and assigning an energy
\begin{equation}
\label{enerprob}
E = \Phi(r_0) - V_c^2 \; {\rm ln}(1 - {\cal R})
\end{equation}
Once the energy is known, the maximum value of $\eta$ corresponding to
that energy can  be calculated using equation~(\ref{etamax}). For  our
random number ${\cal R}$ this yields
\begin{equation}
\label{etamax_random}
\eta_{\rm max} =\sqrt{2 {\rm e}} \; (1-{\cal R}) \;
\sqrt{-{\rm ln}(1-{\cal R})}
\end{equation}
Note that $\eta_{\rm max}$ is independent of the radius $r_0$.

In order to draw $\eta$ from the probability distribution
\begin{equation}
\label{prob_eta}
{\cal  P}(\eta) = {\eta   \;  h(\eta) \over \sqrt{\eta_{\rm  max}^2  -
    \eta^2}}
\end{equation}
we choose the comparison probability distribution
\begin{equation}
\label{prob_comp}
{\cal  P}_{\rm comp}(\eta)  =   {\eta_{\rm max} \over  \sqrt{\eta_{\rm
      max}^2 - \eta^2}}
\end{equation}
which has the property ${\cal P}_{\rm comp}(\eta) \geq {\cal P}(\eta)$
for  $0 \leq \eta \leq  \eta_{\rm max}$, as long  as $h(\eta) \leq 1$. 
Trial values for $\eta$ are  drawn  from the probability  distribution
${\cal P}_{\rm comp}(\eta)$ by inversion of its normalized, cumulative
probability distribution,
\begin{equation}
\label{prob_comp_norm}
\widehat{\cal   P}_{\rm  comp}(  <   \eta)   =  {2  \over   \pi}  {\rm
  arcsin}\left( {\eta \over \eta_{\rm max}} \right),
\end{equation}
and the rejection  method (e.g., Press \etal 1992)  is used  to decide
whether or not this trial value should be accepted.
 
The eccentricity   of  a random orbit    in  an anisotropic,  singular
isothermal sphere is thus obtained as follows:

\begin{enumerate}

\item Draw random numbers ${\cal R}_1$ and ${\cal R}_2$.

\item Calculate  $\eta_{\rm max} = \sqrt{2 {\rm  e}} \; (1-{\cal R}_1)
  \; \sqrt{-{\rm ln}(1-{\cal R}_1)}$. If ${\cal R}_2 > \eta_{\rm max}$
  return to step 1. This takes care of the fact that ${\cal P}(E)$ and
  ${\cal P}(\eta)$ are not   independent: ${\cal P}(\eta)$  depends on
  $\eta_{\rm max}$ which is a function of energy.

\item Draw random numbers ${\cal R}_3$ and ${\cal R}_4$.

\item Calculate $\eta_{\rm try}  = \eta_{\rm max} \sin\bigl({\pi \over
    2} {\cal   R}_3 \bigr)$.     If  ${\cal R}_4   >    {\cal  P}_{\rm
    comp}(\eta_{\rm try}) / {\cal P}(\eta_{\rm try})$ return to step 3

\item   Accept  $\eta_{\rm  try}$  and    compute  the   corresponding
  eccentricity by solving for the roots of equation~(\ref{rootred}).

\end{enumerate}

When determining the  distribution  of  orbital eccentricities  for  a
power-law tracer population in   a singular isothermal potential  (see
Section~\ref{sec:tracers}), one can  use  the same method as  outlined
above but with the equation in step 2 replaced by
\begin{equation}
\label{etamax_random_trace}
\eta_{\rm  max}   =\sqrt{2  {\rm e}}     \; \sqrt{\beta} \;   (1-{\cal
  R}_1)^{\beta} \; \sqrt{-{\rm ln}(1-{\cal R}_1)},
\end{equation}
where $\beta = 1/(\alpha -1)$.

\clearpage
 

\ifsubmode\else
\baselineskip=10pt
\fi


\clearpage


\ifsubmode\else
\baselineskip=14pt
\fi


\newcommand{\figcapeps}{The  eccentricity as  function of  the orbital
  circularity   $\eta$   for   orbits   in   a   singular   isothermal
  sphere.\label{fig:eps}}

\newcommand{\figcapintegral}{The  average eccentricity  of orbits in a
  spherical,   singular  isothermal  potential   as  function  of  the
  anisotropy parameter $a$.  The  discontinuous behavior at $a=0$  is
  due to our particular choice for  the function $h_a(\eta)$. Negative
  and positive  $a$  correspond to  radial  and tangential  anisotropy
  respectively.   In      the isotropic  case    ($a=0$)   the average
  eccentricity is $\overline{e} = 0.55$.\label{fig:integral}}

\newcommand{\figcapanis}{Normalized  distribution  functions  of   the
  eccentricity (upper  panels) and  apo-  to pericenter  ratio  (lower
  panels) of a singular isothermal sphere. Results are shown for three
  values of  the anisotropy parameter  $a$, as indicated  in the upper
  panels. These distributions have been determined using a Monte Carlo
  simulation with   $10^6$ orbits (see  Section~\ref{sec:numer}).  The
  20th, 50th (median), and 80th percentile points of the distributions
  are  indicated   with      dotted,  solid,   and     dashed    lines
  respectively.\label{fig:anis}}

\newcommand{\figcaptrace}{The average  eccentricity as a  function  of
  the power-law   slope  $\alpha$ of the  density   distribution of an
  isotropic tracer population  in   a singular isothermal halo.    The
  distribution  of orbital eccentricities  depends  only mildly on the
  slope of the density distribution.\label{fig:trace}}

\newcommand{\figcaptrunc}{The 20th (dotted lines), 50th (solid lines),
  and   80th   (dashed lines)  percentile    points  of the cumulative
  distributions of  orbital  eccentricities (left panel)   and apo- to
  pericenter ratios (right panel)  as a function  of the ratio of  the
  truncation radius to the   core radius ($r_t/r_c$) of  the truncated
  isothermal  sphere   with a  constant  density   core and  isotropic
  velocity distribution.  The  average orbital eccentricity  increases
  with   increasing  $r_t/r_c$.  In  the  limit  $r_t/r_c  \rightarrow
  \infty$,   the distribution   of   orbital  eccentricities   becomes
  virtually identical to   that  of the  singular  isothermal  sphere. 
  \label{fig:trunc}}

\newcommand{\figcapjafhern}{Comparison of the  normalized eccentricity
  distributions of  three spherical, isotropic potentials:  a singular
  isothermal,  a Hernquist,  and   a Jaffe  sphere.   The eccentricity
  distribution   of  these   three  density  distributions    are only
  moderately different.  The   vertical lines correspond  to the 20th,
  50th, and   80th   percentile points  of  the  distributions  as  in
  Figure~\ref{fig:anis}.\label{fig:jafhern}}

\newcommand{\figcapdensprof}{Density profiles of the dark halo evolved
  in isolation, i.e.,  without satellite.  The dashed lines correspond
  to the analytical profile, whereas the solid  lines show the density
  measured   directly   from  the  particle  distribution  by  binning
  particles in spherical shells with 100 particles each.  The vertical
  dotted lines  correspond  to the core  radius  ($r_c  = 4$  kpc) and
  truncation radius ($r_t  = 200$  kpc)  of the halo.  The upper panel
  shows the initial conditions (at t=0), whereas the lower panel shows
  the density profile at the end of the simulation (at t=10 Gyr).  The
  density profile of the particle  distribution has not changed by any
  significant amount,  suggesting  that the  initial  conditions  were
  already close  to  equilibrium.   We  use  the end result   of  this
  simulation as initial conditions  for   the halo particles for   all
  subsequent runs with orbiting satellites.\label{fig:densprof}}

\newcommand{\figcaporbits}{Orbits of the  satellites in Models~1 to 6. 
  Both  the  $x$--$y$ (upper,  big panels),   and the $x$--$z$ (lower,
  small panels)  projections are shown.  The  solid  dot indicates the
  initial position from which the satellite is started. Parameters for
  the      different         models      are      listed            in
  Table~\ref{tab:param}.\label{fig:orbits}}

\newcommand{\figcapresa}{Galactocentric        distance,       orbital
  eccentricity, energy, and  orbital angular momentum of the satellite
  as function of time for Models~1 to 4.  The panels  in the first row
  show   the galactocentric  distance, i.e, the   distance between the
  satellite and the center-of-mass of  the halo.  Panels in the second
  row show  the orbital eccentricity: the  solid lines show the values
  calculated by solving for the roots of equation~(\ref{rooteq}) using
  the  analytical  potential  of  the  halo,  and   the   open circles
  correspond to the  empirical eccentricity computed using the turning
  points (see Section~\ref{sec:resfric}).  The eccentricities are only
  shown  up to  $t_{0.8}$, since  after  that time  the  satellite  is
  basically just sitting in the   center and the eccentricity  becomes
  meaningless.  Panels   in  the third  row show   the orbital energy,
  normalized by the central value of  the potential $\Phi_0$ at $t=0$,
  and the panels  in the fourth  row show the orbital angular momentum
  of  the     satellite,    normalized   to     its    initial   value
  $L(0)$.\label{fig:resa}}

\newcommand{\figcapresb}{Same as Figure~\ref{fig:resa} except that now
  we plot the results  for Models 5  to 8.  For  Model~6 the orbit  is
  circular and no eccentricities are calculated.\label{fig:resb}}

\newcommand{\figcaphalftime}{The   characteristic dynamical   friction
  times (as defined in Section~\ref{sec:resfric}) as a function of the
  intrinsic  eccentricity of the orbit (Models~3,  4, 5,  and 6).  The
  panel   on  the left   shows  $t_{0.4}$,  $t_{0.6}$, and  $t_{0.8}$,
  describing the  timescales for energy  loss, whereas the right panel
  shows $t_{1/4}$, $t_{1/2}$, and $t_{3/4}$, describing the timescales
  for angular momentum loss.   More eccentric orbits lose their energy
  and  angular momentum more rapidly.  The dotted  line corresponds to
  the predictions from Lacy \& Cole (1993), who calculated a timescale
  behavior of the  form $t \propto  \eta^{0.78}$. Our results imply  a
  somewhat    weaker  dependence;    $t   \propto \eta^{0.53}$    (see
  text).\label{fig:halftime}}

\newcommand{\figcaphalfeps}{Same as   Figure~\ref{fig:halftime} except
  that  we now plot  the  characteristic  times  as functions  of  the
  satellite softening length  $\epsilon_s$  (Models~3, 7,  8, and 9).  
  The dotted line  corresponds to the   predictions for $t_{0.6}$  and
  $t_{1/2}$  scaled  to      those     of Model~3   (see        text). 
  \label{fig:halfeps}}

\newcommand{\figcapcomp}{Comparison of results of Models~4, 10 and~11.
  These   models only  differ  in   the  number of   halo particles as
  indicated in the upper  left panel. As can  be seen, the results are
  very       robust,    with   only      marginal    differences  (see
  text).\label{fig:comp}}

\newcommand{\figcapprob}{Contour plot  of K-S  probabilities  that the
  orbital  eccentricities of the globular clusters  in the OBGT sample
  are drawn  randomly from  the expected  distribution of  a system of
  globular clusters with a power-law  density distribution with  slope
  $\alpha$     and       anisotropy        parameter    $a$       (see
  equation~[\ref{anisotropy}])  embedded   in   a  singular isothermal
  sphere.  The  contours correspond  to the $10,20,...,80,90$  percent
  probability  levels,  whereby the   latter is  plotted  as a   thick
  contour.   Note   that  the orbital  eccentricities  contain  little
  information about the   actual density distribution of  the globular
  cluster system.     However,   the  velocity  anisotropy   is   well
  constrained, and a small amount of radial  anisotropy is required in
  order to  explain  the  observed  eccentricities  of   the  globular
  clusters.  The solid dot corresponds to the  best fitting model with
  $\alpha     =  3.5$, the  results    for    which  are plotted    in
  Figure~\ref{fig:ksplot}. \label{fig:prob}}

\newcommand{\figcapksplot}{The   thin line     shows   the  cumulative
  distribution function  of orbital eccentricities of   a sample of 15
  globular clusters in the Milky Way halo studied by Odenkirchen \etal
  (1997).    The thick solid lines    corresponds  to the   cumulative
  distribution of orbital   eccentricities of the  best fitting tracer
  population  with a  $r^{-3.5}$ density   distribution embedded in  a
  singular   isothermal   sphere,   determined  using  a  Monte  Carlo
  simulation  as described  in  Section~\ref{sec:tracers}  with $10^6$
  orbits.  This model  has a  mild  radial velocity anisotropy and  is
  indicated by a solid  dot in Figure~\ref{fig:prob}.  An  application
  of the K-S test indicates that there is a $94.3$ percent probability
  that  both data   sets   are drawn from    the same  distribution.   
  \label{fig:ksplot}}

\newcommand{\figcapghigna}{The   thin   line    shows  the  cumulative
  distribution function of orbital eccentricities of  the sample of 98
  subhalos in the high  resolution $N$-body cluster analyzed by Ghigna
  \etal (1998) that have apocenters within the cluster's virialization
  radius.  The thick solid    lines  corresponds to  the    cumulative
  distribution  of orbital eccentricities of   the best fitting tracer
  population  with a  $r^{-2.0}$ density  distribution  embedded in  a
  singular isothermal   sphere.  This  model has  a  very  mild radial
  velocity anisotropy  of $a = -0.04$. An  application of the K-S test
  indicates that there is a $61.3$ percent  probability that both data
  sets are drawn from the same distribution.
  \label{fig:ghigna}}

\newcommand{\figcapavertime}{The    average   dynamical friction  time
  $\langle t/t_0 \rangle$,   normalized to that  of a  circular orbit,
  $t_0$, of  a   singular isothermal  sphere  as  a  function  of  the
  anisotropy parameter $a$ (see equation~[\ref{frictime}]).  The solid
  line corresponds   to $t \propto  \eta^{0.53}$, consistent  with the
  results from  our  numerical  simulations, whereas the  dashed  line
  corresponds to  $t \propto  \eta^{0.78}$,  as predicted by  Lacey \&
  Cole (1993). \label{fig:avertime}}


\ifsubmode
\figcaption{\figcapeps}
\figcaption{\figcapintegral}
\figcaption{\figcapanis}
\figcaption{\figcaptrace}
\figcaption{\figcaptrunc}
\figcaption{\figcapjafhern}
\figcaption{\figcapdensprof}
\figcaption{\figcaporbits}
\figcaption{\figcapresa}
\figcaption{\figcapresb}
\figcaption{\figcaphalftime}
\figcaption{\figcaphalfeps}
\figcaption{\figcapcomp}
\figcaption{\figcapprob}
\figcaption{\figcapksplot}
\figcaption{\figcapghigna}
\figcaption{\figcapavertime}
\clearpage
\else\printfigtrue\fi

\ifprintfig

\clearpage

\clearpage
\begin{figure}
\epsfxsize=7.0truecm
\centerline{\epsfbox{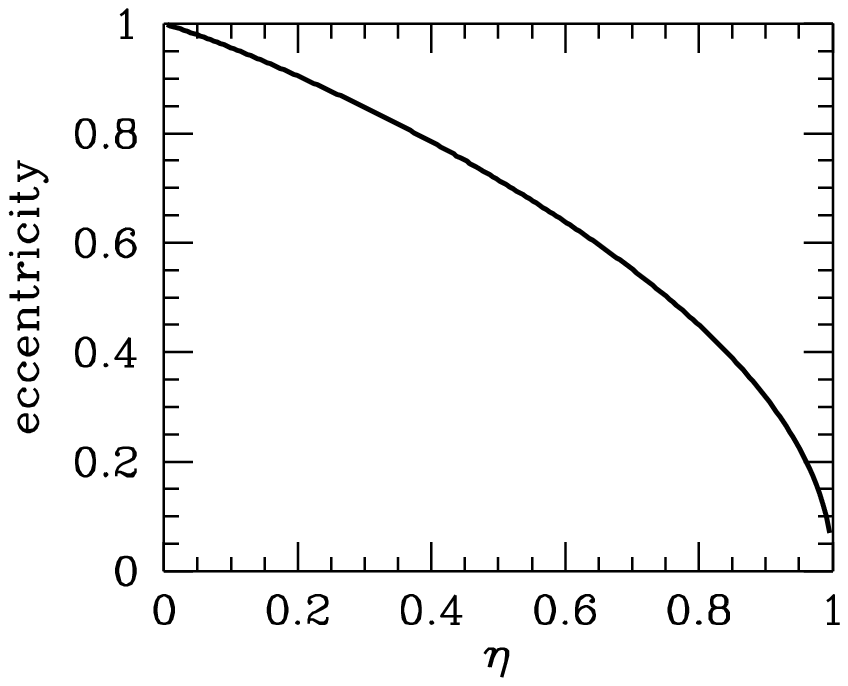}}
\ifsubmode
\vskip3.0truecm
\setcounter{figure}{0}
\addtocounter{figure}{1}
\centerline{Figure~\thefigure}
\else\figcaption{\figcapeps}\fi
\end{figure}


\clearpage
\begin{figure}
\epsfxsize=7.0truecm
\centerline{\epsfbox{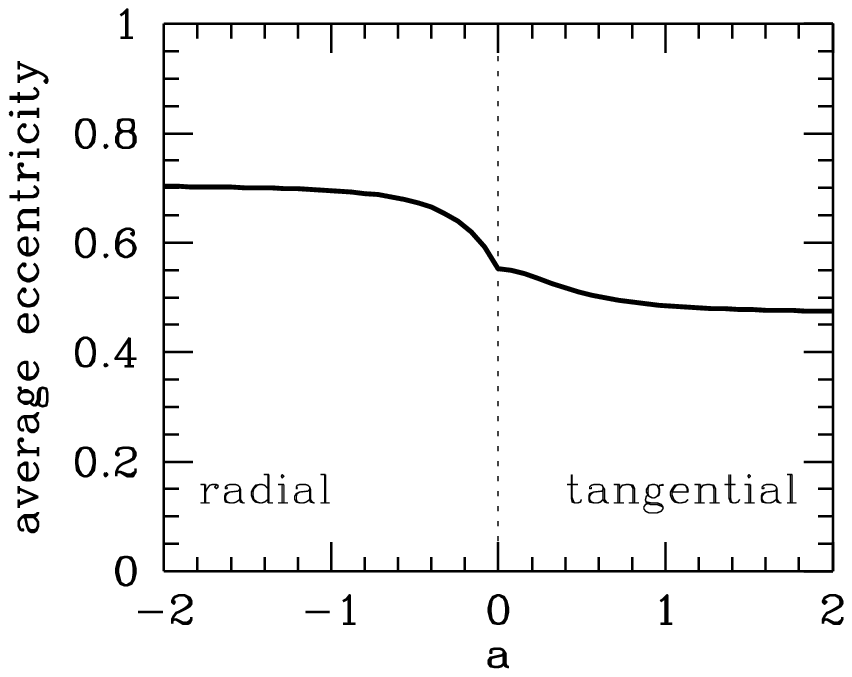}}
\ifsubmode
\vskip3.0truecm
\addtocounter{figure}{1}
\centerline{Figure~\thefigure}
\else\figcaption{\figcapintegral}\fi
\end{figure}


\clearpage
\begin{figure}
\epsfxsize=16.0truecm
\centerline{\epsfbox{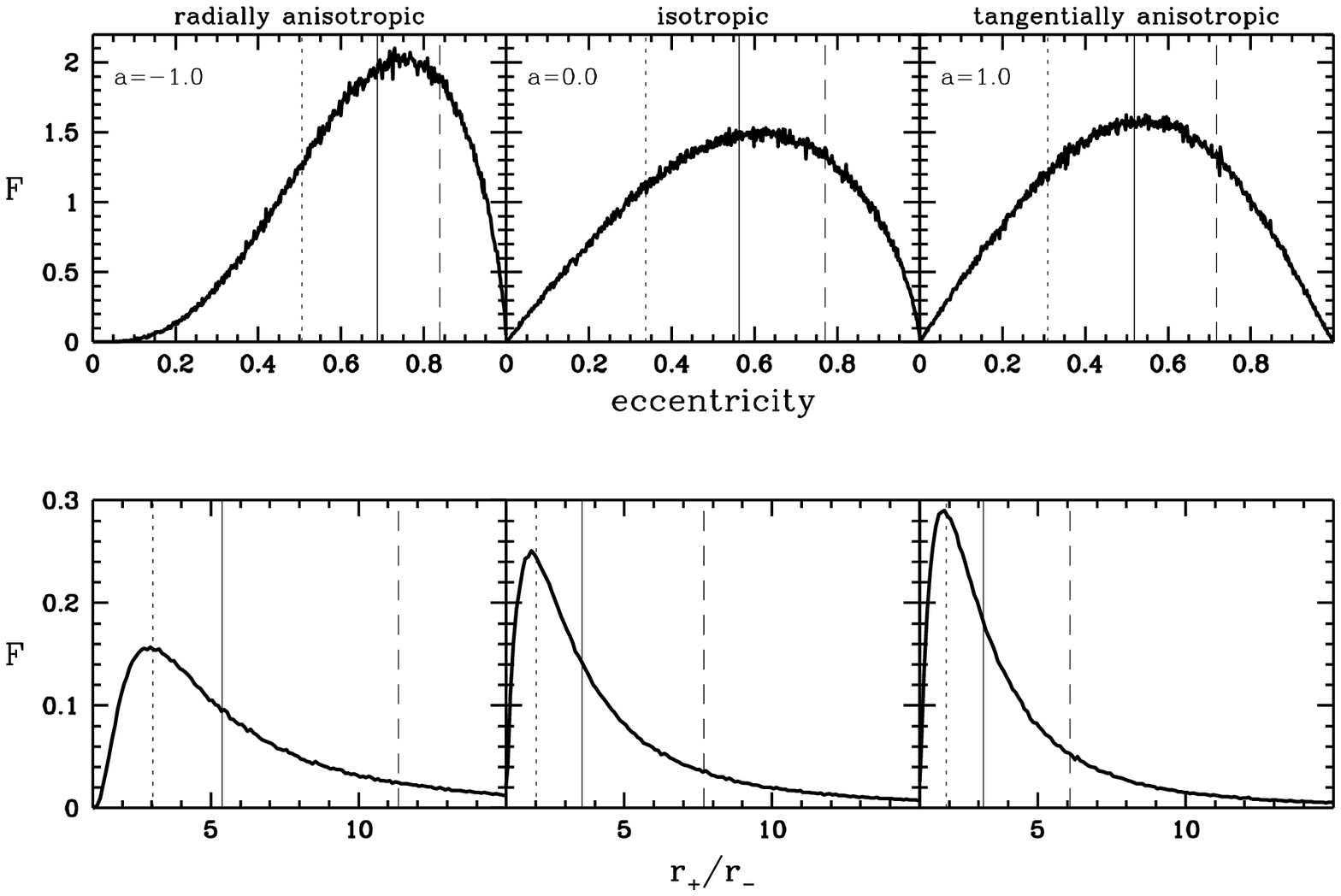}}
\ifsubmode
\vskip3.0truecm
\addtocounter{figure}{1}
\centerline{Figure~\thefigure}
\else\figcaption{\figcapanis}\fi
\end{figure}


\clearpage
\begin{figure}
\epsfxsize=7.0truecm
\centerline{\epsfbox{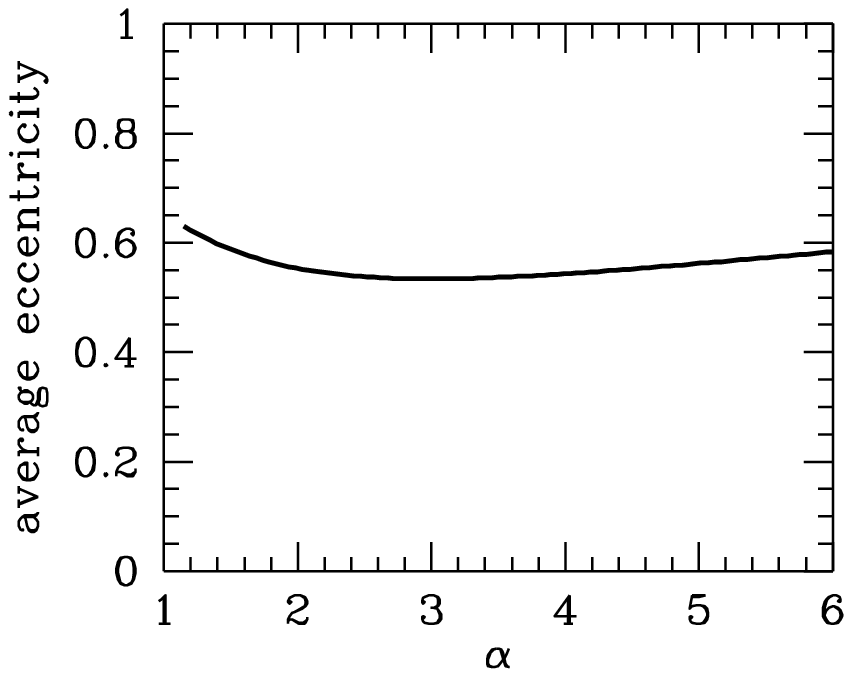}}
\ifsubmode
\vskip3.0truecm
\addtocounter{figure}{1}
\centerline{Figure~\thefigure}
\else\figcaption{\figcaptrace}\fi
\end{figure}


\clearpage
\begin{figure}
\epsfxsize=16.0truecm
\centerline{\epsfbox{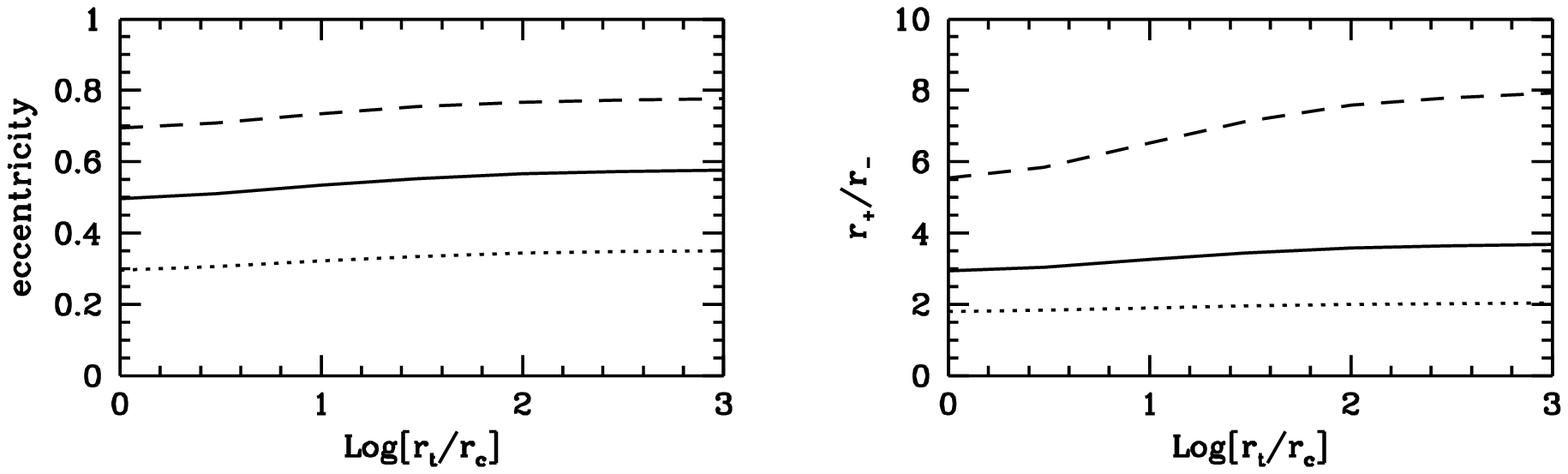}}
\ifsubmode
\vskip3.0truecm
\addtocounter{figure}{1}
\centerline{Figure~\thefigure}
\else\figcaption{\figcaptrunc}\fi
\end{figure}


\clearpage
\begin{figure}
\epsfxsize=7.0truecm
\centerline{\epsfbox{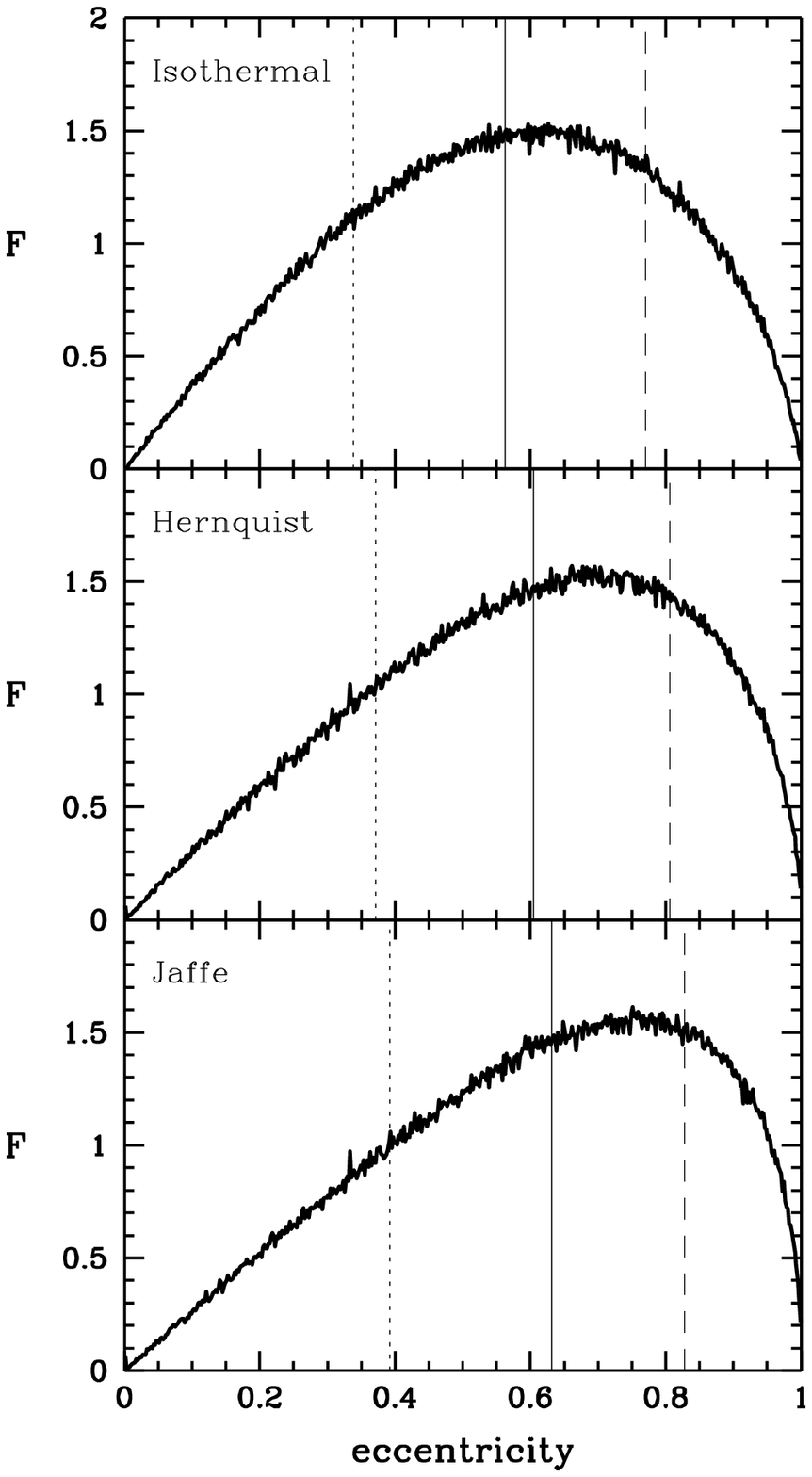}}
\ifsubmode
\vskip3.0truecm
\addtocounter{figure}{1}
\centerline{Figure~\thefigure}
\else\figcaption{\figcapjafhern}\fi
\end{figure}


\clearpage
\begin{figure}
\epsfxsize=7.0truecm
\centerline{\epsfbox{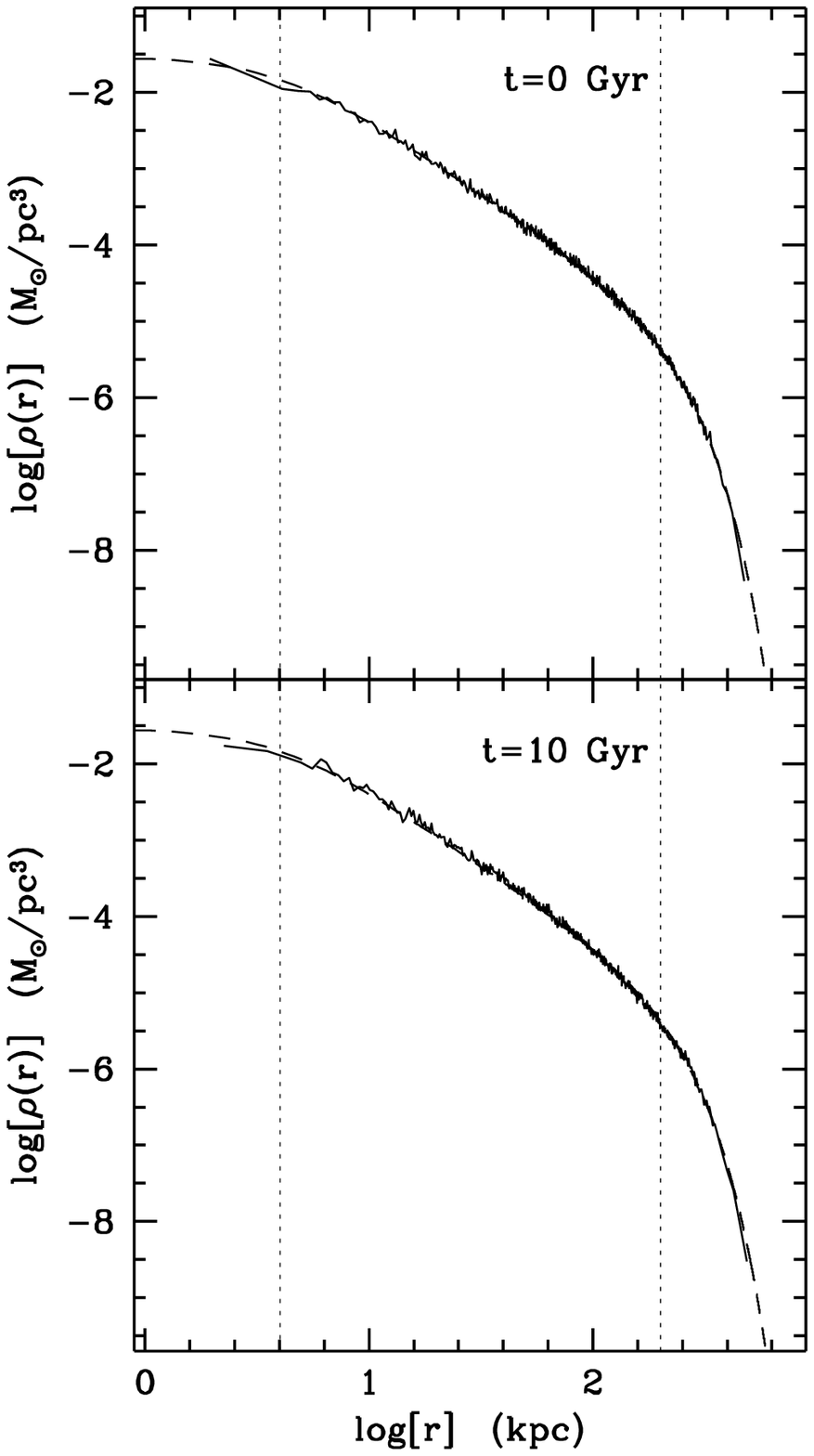}}
\ifsubmode
\vskip3.0truecm
\addtocounter{figure}{1}
\centerline{Figure~\thefigure}
\else\figcaption{\figcapdensprof}\fi
\end{figure}


\clearpage
\begin{figure}
\epsfxsize=16.0truecm
\centerline{\epsfbox{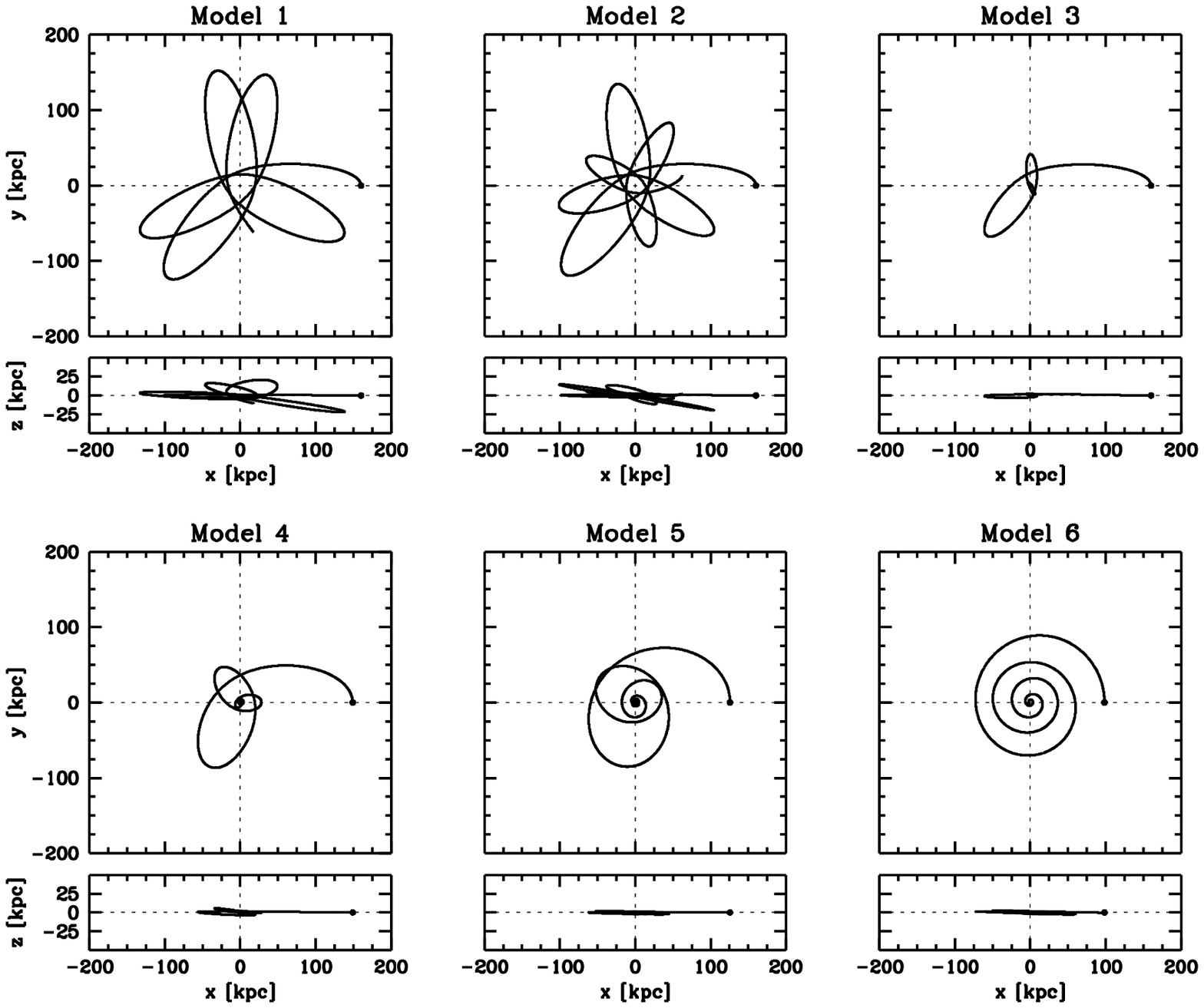}}
\ifsubmode
\vskip3.0truecm
\addtocounter{figure}{1}
\centerline{Figure~\thefigure}
\else\figcaption{\figcaporbits}\fi
\end{figure}


\clearpage
\begin{figure}
\epsfxsize=16.0truecm
\centerline{\epsfbox{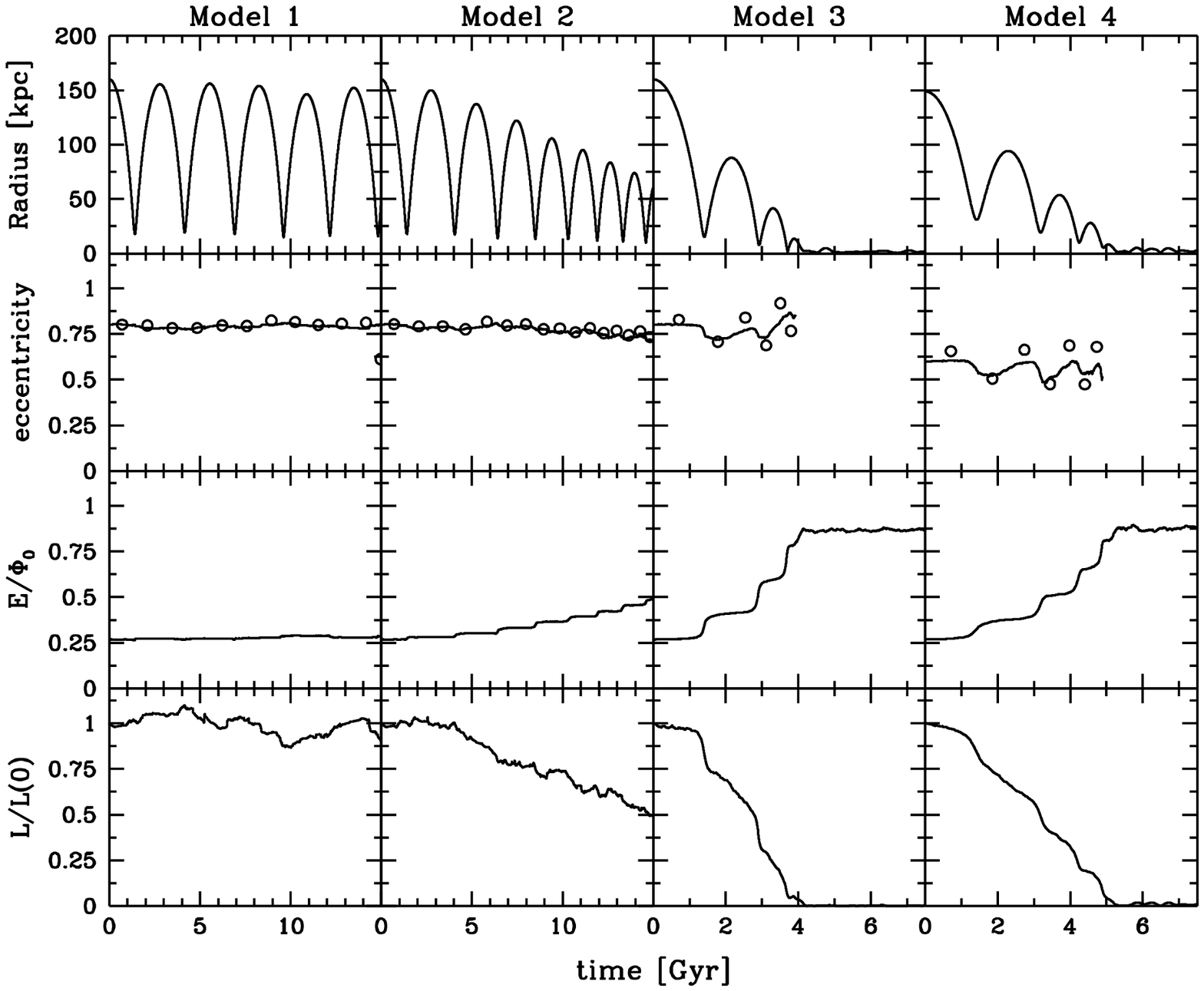}}
\ifsubmode
\vskip3.0truecm
\addtocounter{figure}{1}
\centerline{Figure~\thefigure}
\else\figcaption{\figcapresa}\fi
\end{figure}


\clearpage
\begin{figure}
\epsfxsize=16.0truecm
\centerline{\epsfbox{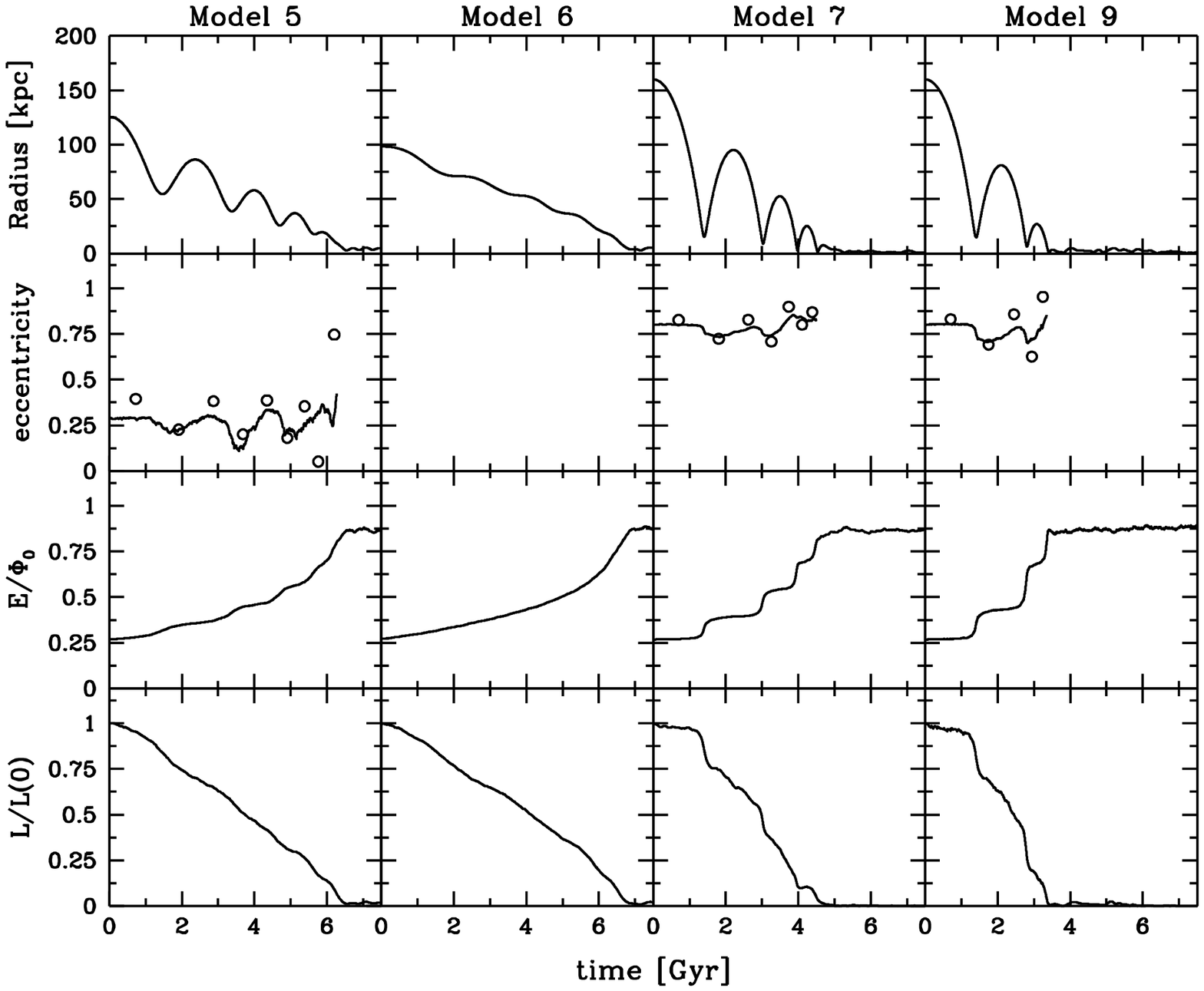}}
\ifsubmode
\vskip3.0truecm
\addtocounter{figure}{1}
\centerline{Figure~\thefigure}
\else\figcaption{\figcapresb}\fi
\end{figure}


\clearpage
\begin{figure}
\epsfxsize=16.0truecm
\centerline{\epsfbox{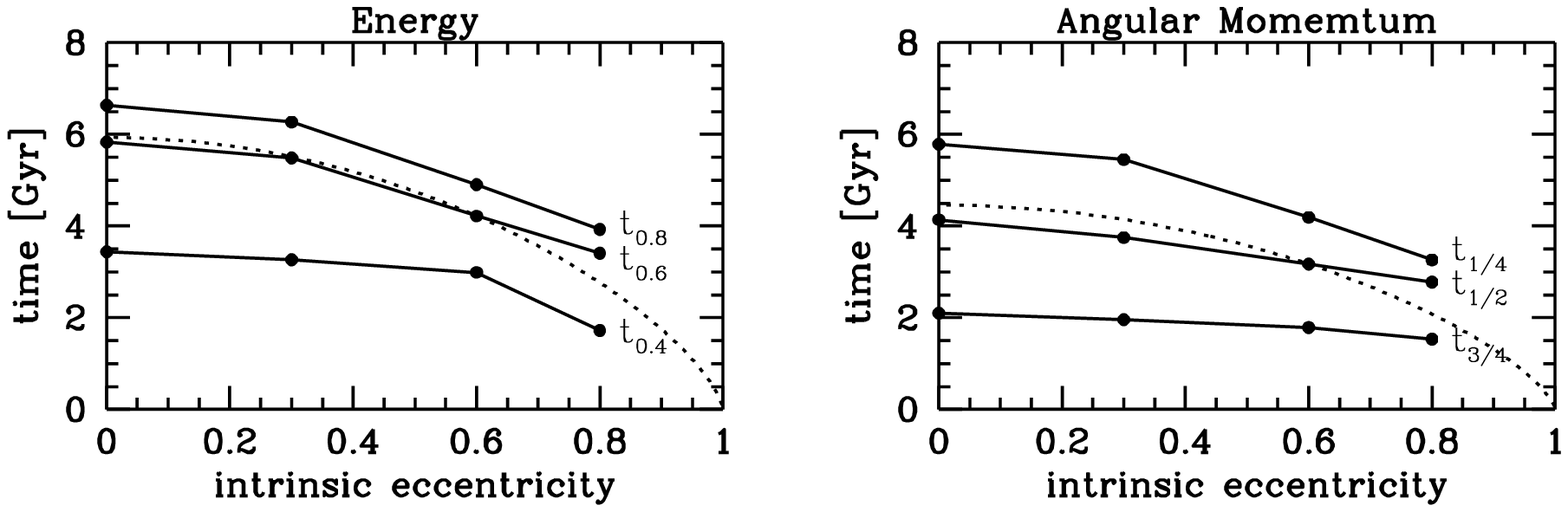}}
\ifsubmode
\vskip3.0truecm
\addtocounter{figure}{1}
\centerline{Figure~\thefigure}
\else\figcaption{\figcaphalftime}\fi
\end{figure}


\clearpage
\begin{figure}
\epsfxsize=16.0truecm
\centerline{\epsfbox{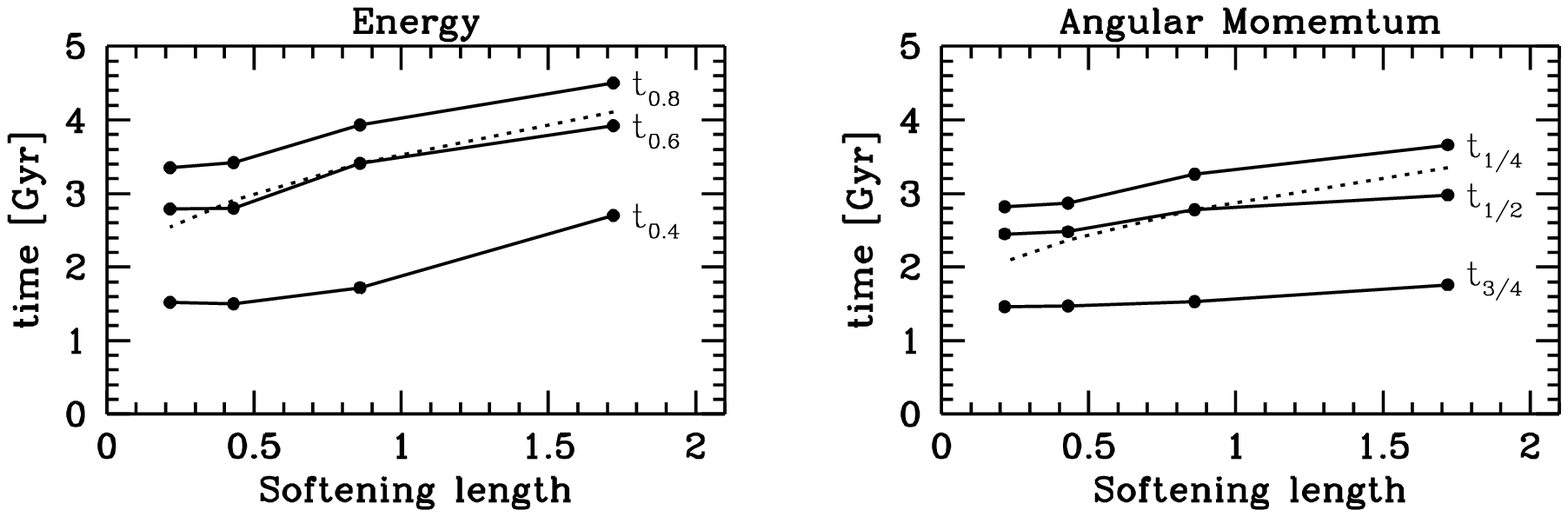}}
\ifsubmode
\vskip3.0truecm
\addtocounter{figure}{1}
\centerline{Figure~\thefigure}
\else\figcaption{\figcaphalfeps}\fi
\end{figure}


\clearpage
\begin{figure}
\epsfxsize=16.0truecm
\centerline{\epsfbox{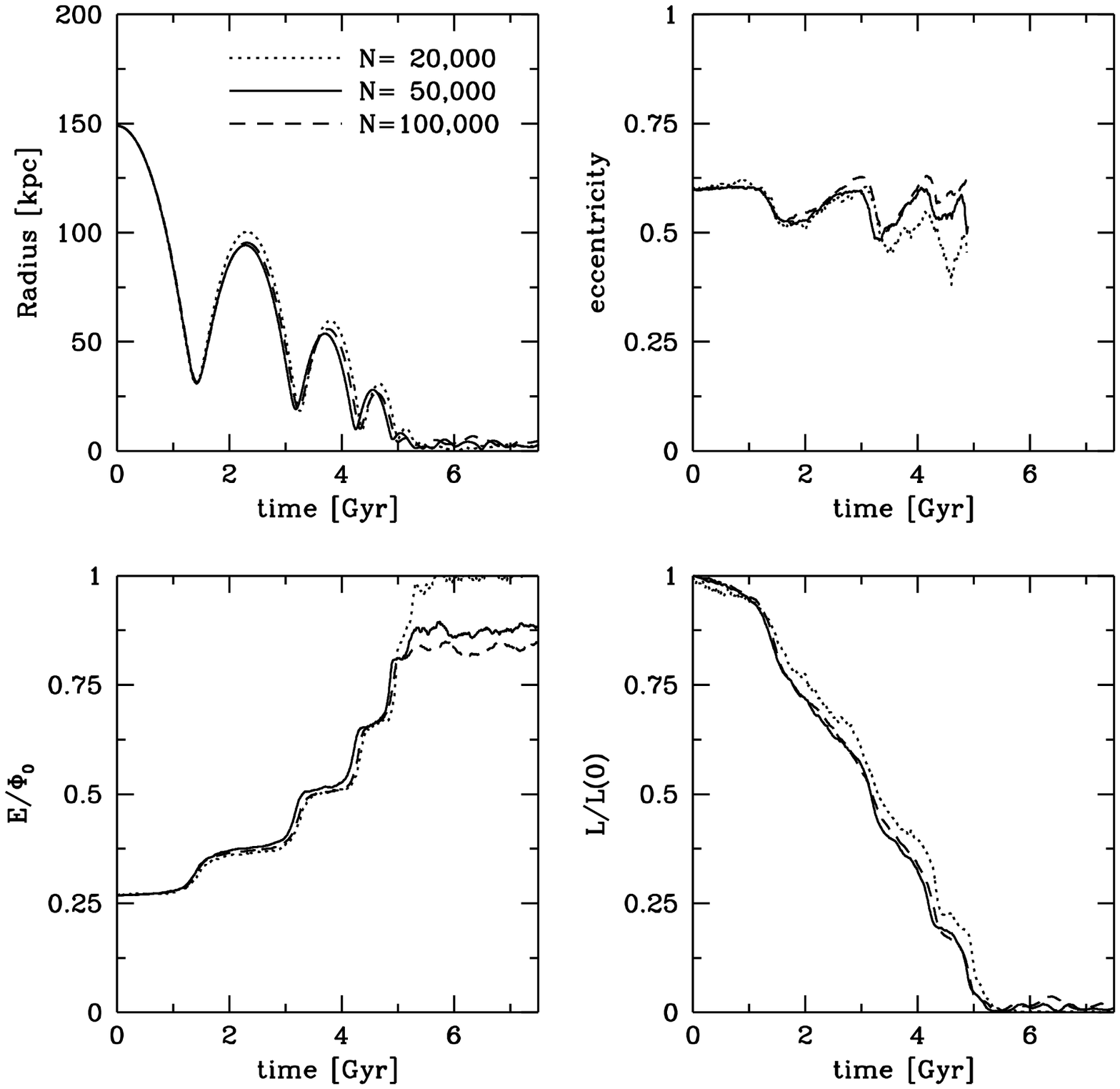}}
\ifsubmode
\vskip3.0truecm
\addtocounter{figure}{1}
\centerline{Figure~\thefigure}
\else\figcaption{\figcapcomp}\fi
\end{figure}


\clearpage
\begin{figure}
\epsfxsize=10.0truecm
\centerline{\epsfbox{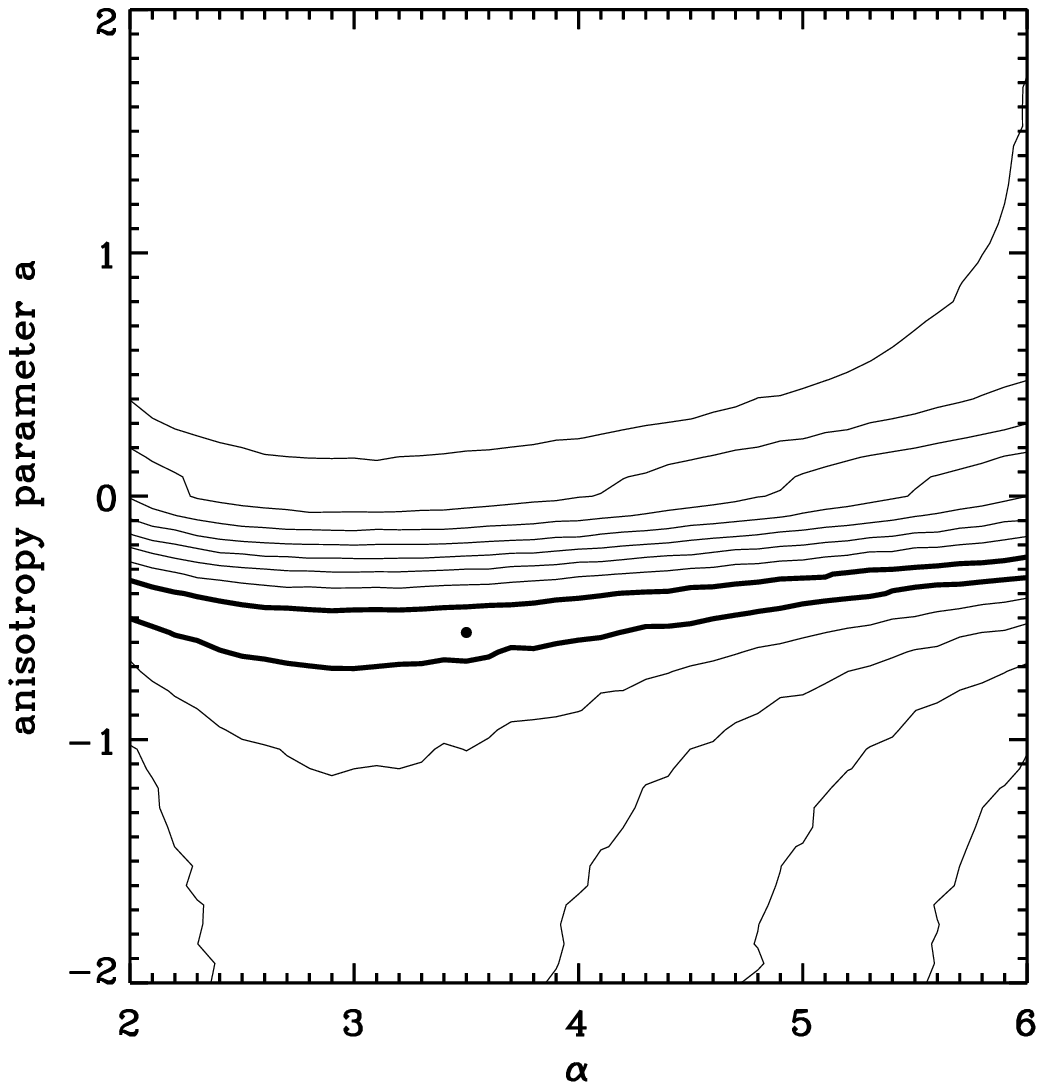}}
\ifsubmode
\vskip3.0truecm
\addtocounter{figure}{1}
\centerline{Figure~\thefigure}
\else\figcaption{\figcapprob}\fi
\end{figure}


\clearpage
\begin{figure}
\epsfxsize=10.0truecm
\centerline{\epsfbox{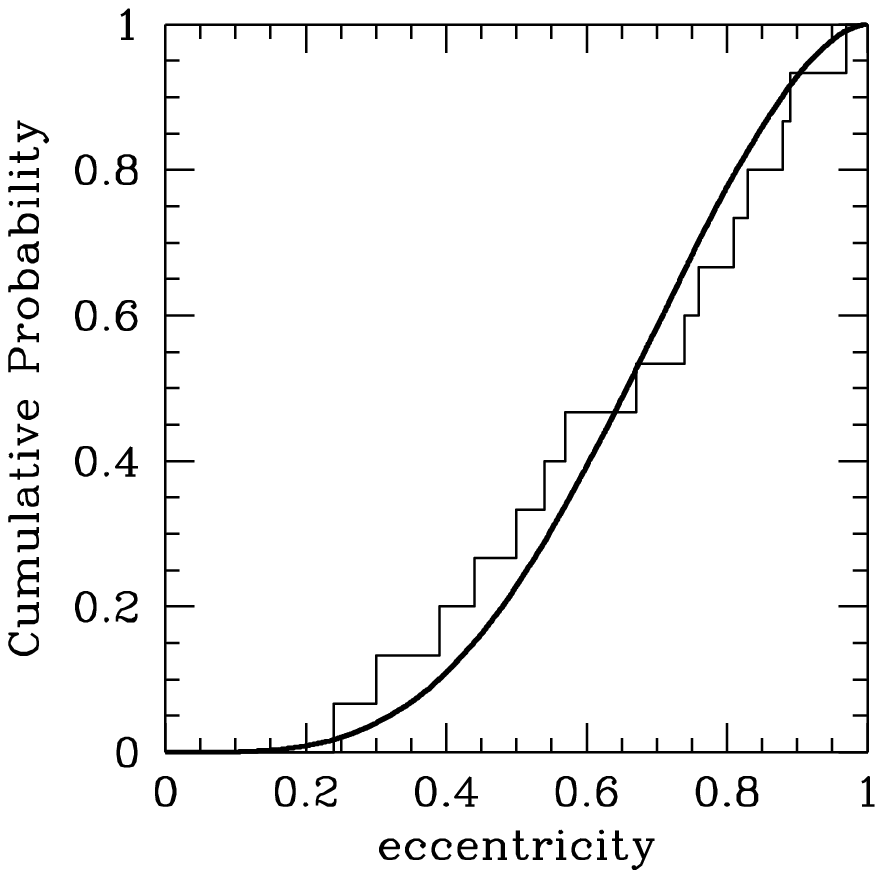}}
\ifsubmode
\vskip3.0truecm
\addtocounter{figure}{1}
\centerline{Figure~\thefigure}
\else\figcaption{\figcapksplot}\fi
\end{figure}


\clearpage
\begin{figure}
\epsfxsize=10.0truecm
\centerline{\epsfbox{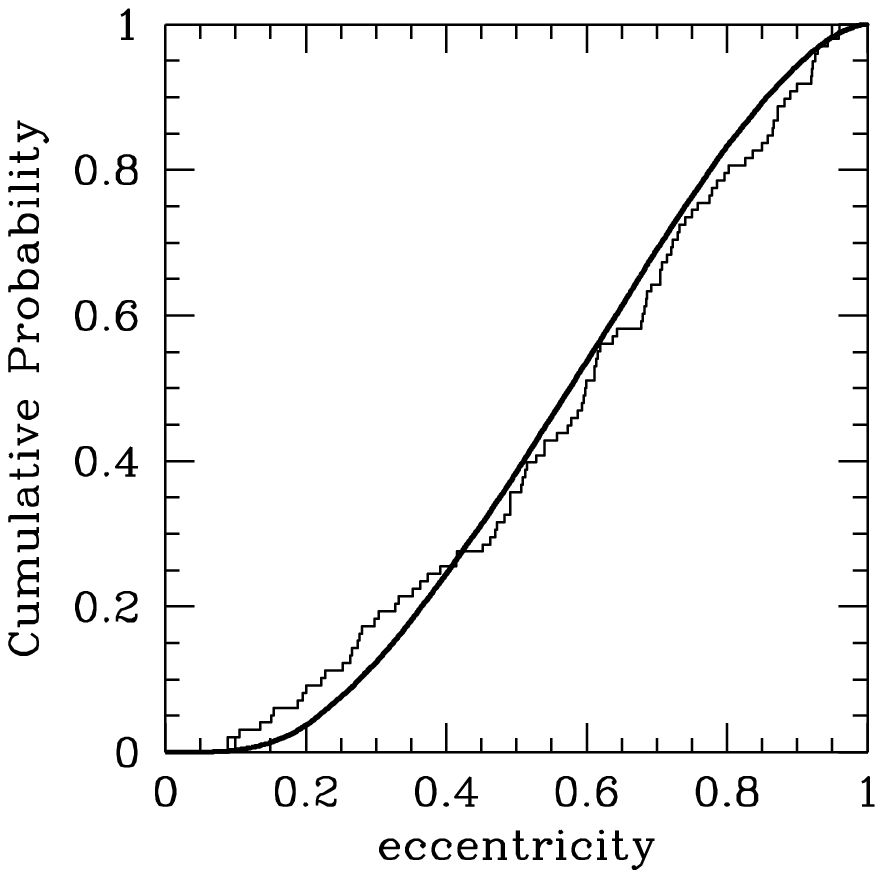}}
\ifsubmode
\vskip3.0truecm
\addtocounter{figure}{1}
\centerline{Figure~\thefigure}
\else\figcaption{\figcapghigna}\fi
\end{figure}


\clearpage
\begin{figure}
\epsfxsize=7.0truecm
\centerline{\epsfbox{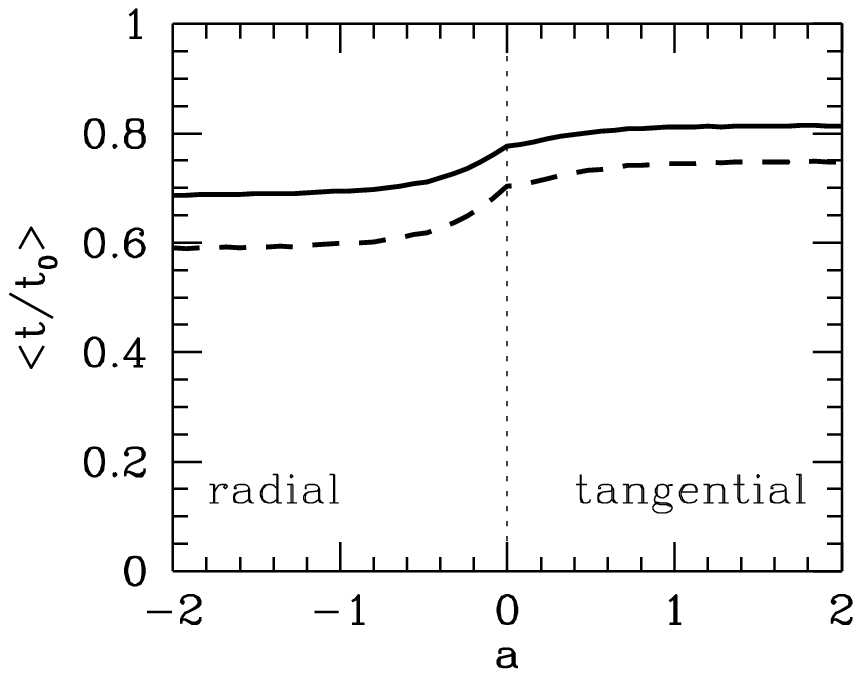}}
\ifsubmode
\vskip3.0truecm
\addtocounter{figure}{1}
\centerline{Figure~\thefigure}
\else\figcaption{\figcapavertime}\fi
\end{figure}

\fi



\clearpage
\ifsubmode\pagestyle{empty}\fi


\begin{deluxetable}{rccccc}
\tablecaption{Parameters of $N$-body simulations \label{tab:param}}
\tablehead{
\colhead{Model} & \colhead{$M_s$} & \colhead{$e_0$} & 
\colhead{$r_{+}$} & \colhead{$\epsilon_s$} & \colhead{$N_h$}  \\ 
\colhead{(1)} & \colhead{(2)} & \colhead{(3)} & 
\colhead{(4)} & \colhead{(5)} & \colhead{(6)} \\
}
\startdata
1 & $2\times 10^{-4}$ & $0.8$ & $40.0$ & $0.18$ & $5 \times 10^4$ \\
2 & $2\times 10^{-3}$ & $0.8$ & $40.0$ & $0.40$ & $5 \times 10^4$ \\
3 & $2\times 10^{-2}$ & $0.8$ & $40.0$ & $0.86$ & $5 \times 10^4$ \\
4 & $2\times 10^{-2}$ & $0.6$ & $37.3$ & $0.86$ & $5 \times 10^4$ \\
5 & $2\times 10^{-2}$ & $0.3$ & $31.4$ & $0.86$ & $5 \times 10^4$ \\
6 & $2\times 10^{-2}$ & $0.0$ & $24.6$ & $0.86$ & $5 \times 10^4$ \\
7 & $2\times 10^{-2}$ & $0.8$ & $40.0$ & $1.72$ & $5 \times 10^4$ \\
8 & $2\times 10^{-2}$ & $0.8$ & $40.0$ & $0.43$ & $5 \times 10^4$ \\
9 & $2\times 10^{-2}$ & $0.8$ & $40.0$ & $0.22$ & $5 \times 10^4$ \\
10 & $2\times 10^{-2}$ & $0.6$ & $37.3$ & $0.86$ & $2 \times 10^4$ \\
11 & $2\times 10^{-2}$ & $0.6$ & $37.3$ & $0.86$ & $1 \times 10^5$ \\
\enddata

\tablecomments{Column~(1) lists  the model ID  by which  the different
  simulations are addressed  in the text.   Columns~(2), (3), (4), and
  (5) list  the  masses, the   initial eccentricities, the  apocentric
  radii   from which  the satellites   are  started, and the softening
  lengths of the satellites in each  model, respectively (all in model
  units). Finally, column~(6) lists the  number of halo particles used
  in  the simulations.  All satellites  have the same initial specific
  energy. The satellites in Models 1 to  6 and 10 and  11 all have the
  same average density.}
\end{deluxetable}

\clearpage


\begin{deluxetable}{rrrrrrr}
\tablecaption{Dynamical friction timescales. \label{tab:timescale}}
\tablehead{
\colhead{Model} & \colhead{$t_{0.4}$} & \colhead{$t_{0.6}$} & 
\colhead{$t_{0.8}$} & \colhead{$t_{3/4}$} & 
\colhead{$t_{1/2}$} & \colhead{$t_{1/4}$} \\ 
\colhead{(1)} & \colhead{(2)} & \colhead{(3)} & 
\colhead{(4)} & \colhead{(5)} & \colhead{(6)} & \colhead{(7)} \\
}
\startdata
1 & $\gg 15$ & $\gg 15$ & $\gg 15$ & $\gg 15$   & $\gg 15$  & $\gg 15$\\
2 & $11.88$  & $> 15$   & $\gg 15$ & $\sim 8.5$ & $\sim 15$ & $> 15$  \\
3 & $1.72$   & $3.41$   & $3.93$   & $1.53$     & $2.78$    & $3.26$  \\ 
4 & $2.99$   & $4.22$   & $4.90$   & $1.79$     & $3.17$    & $4.19$  \\
5 & $3.27$   & $5.49$   & $6.27$   & $1.96$     & $3.75$    & $5.45$  \\
6 & $3.44$   & $5.83$   & $6.64$   & $2.11$     & $4.13$    & $5.79$  \\
7 & $2.70$   & $3.92$   & $4.50$   & $1.76$     & $2.98$    & $3.66$  \\
8 & $1.50$   & $2.80$   & $3.42$   & $1.47$     & $2.48$    & $2.87$  \\
9 & $1.52$   & $2.79$   & $3.35$   & $1.46$     & $2.45$    & $2.82$  \\
10 & $3.11$   & $4.35$   & $4.97$   & $2.05$     & $3.29$    & $4.37$  \\
11 & $3.08$   & $4.31$   & $4.98$   & $1.78$     & $3.19$    & $4.27$  \\
\enddata

\tablecomments{The   characteristic timescales in   Gyrs for dynamical
  friction as  defined in Section~\ref{sec:resfric}.  Column~(1) lists
  the model  ID. Columns~(2),  (3),  and (4)  list  the characteristic
  timescales  for energy loss, and Columns~(5),  (6), and (7) list the
  characteristic timescales for angular momentum loss.}
\end{deluxetable}


\begin{thebibliography}{}

\bibitem[]{All91} 
Allen, C., \& Santillan, A. 1991, Revista Mexicana de Astronomia y 
Astrofisica, 22, 255

\bibitem[]{Bau96}
Baugh, C. M., Cole, S., \& Frenk, C. S. 1996, \mnras , 283, 1361

\bibitem[]{Bin87} 
Binney, J. J., \& Tremaine, S. D. 1987, Galactic Dynamics. 
(Princeton: Princeton University Press)

\bibitem[]{Bon91}
Bond, J. R., Cole, S., Efstathiou, G., Kaiser, N., 1991, \apj , 379,
  440

\bibitem[]{Bon87}
Bontekoe, Tj. R., \& van Albada, T. S. 1987, \mnras , 224, 349

\bibitem[]{Bow91}
Bower, R., 1991, \mnras , 248, 332

\bibitem[]{Bra97}
Brainerd, T. G., Goldberg, D. M., \& Villumsen, J. V. 1998, \apj ,
  502, 505

\bibitem[]{Cha43}
Chandrasekhar, S. 1943, \apj , 97, 255

\bibitem[]{Col94}
Cole, S., Arag\'on-Salamanca, A., Frenk, C. S., Navarro, J. F., \&
  Zepf, S. E. 1994, \mnras , 271, 781

\bibitem[]{Col98}
Colpi, M. 1998, \apj , 502, 167

\bibitem[]{Cor97}
Cora, S. A., Muzzio, J. A., \& Vergne, M. M. 1997, \mnras , 289 ,253

\bibitem[]{deV72}
de Vaucouleurs, G., \& Freeman, K. C. 1972, Vistas in Astronomy,
  14, ed. A. Beer (Oxford: Pergamon), p. 163

\bibitem[]{Dik96}
Dikaiakos, M. \& Stadel, J. 1996, ICS Conference Proceedings 1996

\bibitem[]{Ger91}
Gerhard, O. E. 1991, \mnras, 250, 812

\bibitem[]{Ghi98}
Ghigna, S., Moore, B., Governato, F., Lake, G., Quinn, T., \& Stadel,
  J. 1998, \mnras , 300, 146

\bibitem[]{Har76}
Harris, W. E. 1976, \aj , 81, 1095

\bibitem[]{Hey95}
Heyl, J. S., Cole, S., Frenk, C. S., \& Navarro, J. F. 1995, \mnras ,
  274 , 755

\bibitem[]{Her90}
Hernquist, L. 1990, \apj , 356, 359

\bibitem[]{Her93}
Hernquist, L. 1993, \apjs , 86, 389

\bibitem[]{HK89}
Hernquist, L., \& Katz, N. 1989, \apjs , 70, 419

\bibitem[]{HW89}
Hernquist, L., \& Weinberg, M. D. 1989, \mnras , 238, 407

\bibitem[]{Hua97}
Huang, S., \& Carlberg, R. G. 1997, \apj , 480, 503

\bibitem[]{Iba98}
Ibata, R. A., \& Lewis, G. F. 1998, \apj , 500, 575

\bibitem[]{Jaf83}
Jaffe, W. 1983, \mnras , 202, 995

\bibitem[]{Joh95}
Johnston, K. V., Spergel, D. N., \& Hernquist, L. 1995, \apj , 451, 598

\bibitem[]{Joh96}
Johnston, K. V., Hernquist, L., \& Bolte, M. 1996, \apj , 465, 278

\bibitem[]{Joh98}
Johnston, K. V. 1998, \apj , 297, 308

\bibitem[]{Kau93}
Kauffmann, G., White, S. D. M., \& Guiderdoni, B. 1993, \mnras , 264,
201
 
\bibitem[]{Kin62}
King, I. R. 1962, \aj , 67, 471
 
\bibitem[]{Kle98}
Klessen, R. S., \& Kroupa, P. 1998, \apj , 498, 143

\bibitem[]{Kly97}
Klypin, A., Gottl\"ober, S., Kravtsov, A., \& Khokhlov, A. M. 1997,
  preprint (astro-ph/9708191)

\bibitem[]{Kro97}
Kroupa, P. 1997, New Astronomy, 2, 139

\bibitem[]{Lac93}
Lacey, C., \& Cole, S. 1993, \mnras , 262, 627

\bibitem[]{Lin83}
Lin, D. N. C., \& Tremaine, S. 1983, \apj , 264, 364

\bibitem[]{McG90}
McGlynn, T. A., 1990, \apj , 348, 515

\bibitem[]{Moo94}
Moore, B., \& Davis, M. 1994, \mnras , 270, 209

\bibitem[]{Mo96a}
Moore, B., Katz, N., Lake, G., Dressler, A., \& Oemler, A. Jr. 1996,
  \nat , 379, 613

\bibitem[]{Mo96b}
Moore, B., Katz, N., Lake, G. 1996, \apj , 457, 455

\bibitem[]{Mo98a}
Moore, B., Lake, G., \& Katz, N. 1998, \apj , 495, 139

\bibitem[]{Mo98b}
Moore, B., Governato, F., Quinn, T., Stadel, J., \& Lake, G. 1998,
   \apj , 499, 5

\bibitem[]{Mor89}
Morgan, S. \& Lake, G. 1989, \apj , 339, 171.

\bibitem[]{Nav95} 
Navarro, J. F., Frenk, C. S., \& White, S. D. M. 1995, \mnras , 275, 720

\bibitem[]{Nav96} 
Navarro, J. F., Frenk, C. S., \& White, S. D. M. 1996, \apj , 462, 563

\bibitem[]{Nav97} 
Navarro, J. F., Frenk, C. S., \& White, S. D. M. 1997, \apj , 490, 493

\bibitem[]{Nor97} 
Norris, J. E., Freeman, K. C., Mayor, M. and Seitzer, P. 1997, \apjl , 487, L187.

\bibitem[]{Ode97}
Odenkirchen, M., Brosche, P., Geffert, M., \& Tucholke, H.-J. 1997,
  New Astronomy, 2, 477

\bibitem[]{OhK95}
Oh, K. S., Lin, D. N. C., \& Aarseth, S. J. 1995, \apj , 442, 142

\bibitem[]{Pet94}
Peterson, R. C. \& Cudworth, K. 1994, \apj , 420, 612

\bibitem[]{Pia95}
Piatek, S., \& Pryor, C. 1995, \aj , 109, 1071

\bibitem[]{Pre92}
Press, W. H., Teukolsky, S. A., Vetterling, W. T., \& Flannery,
  B. P. 1992, Numerical Recipes (Cambridge: Cambridge University Press)

\bibitem[]{Qui86}
Quinn, P. J., \& Goodman, J. 1986, \apj , 309, 472

\bibitem[]{Qui93}
Quinn, P. J., \& Hernquist, L., \& Fullager, D. P. 1993, \apj , 403,
  74

\bibitem[]{Qui97}
Quinn, T., Katz, N., Stadel, J., \& Lake, G. 1997, preprint (astro-ph/9710043)

\bibitem[]{Sch92}
Schommer, R. A., Olszewski, E. W., Suntzeff, N. B., \& Harris,
  H. C. 1992, \aj , 103, 447

\bibitem[]{Som98}
Somerville, R. S., \& Primack, J. R. 1998, preprint (astro-ph/9802268)

\bibitem[]{Sta98}
Stadel, J., \& Quinn T. 1998, in preparation
 
\bibitem[]{Tot92}
T\'oth, G., \& Ostriker, J. P. 1992, \apj , 389, 5

\bibitem[]{Tre76} 
Tremaine, S. D. 1976, \apj , 203, 72

\bibitem[]{Tre81} 
Tremaine, S. D. 1981, in {\it The Structure and Evolution of Normal
  Galaxies}, eds. S. M. Fall and D. Lynden-Bell (Cambridge: Cambridge
  University Press), p. 67

\bibitem[]{Tre84} 
Tremaine, S. D., \& Weinberg, M. D. 1984, \mnras , 209, 729

\bibitem[]{van97}
van der Marel, R. P., Sigurdsson, S., \& Hernquist, L. 1997, \apj,
  487, 153

\bibitem[]{Wal96}
Walker, I. R., Mihos, C., \& Hernquist, L. 1996, \apj , 460, 121

\bibitem[]{Whi76a}
White, S. D. M. 1976a, \mnras , 174, 19

\bibitem[]{Whi76b}
White, S. D. M. 1976b, \mnras , 174, 467

\bibitem[]{Whi78}
White, S. D. M. 1978, \mnras , 184, 185

\bibitem[]{Whi83}
White, S. D. M. 1983, \apj , 274, 53

\bibitem[]{WhR78}
White, S. D. M., \& Rees, M. J. 1978, \mnras , 183, 341

\bibitem[]{Zar88}
Zaritsky, D., \& White, S. D. M. 1988, \mnras , 235, 289

\bibitem[]{Zha98}
Zhao, H. S. 1998, \mnras, 294, 139

\bibitem[]{zin85}
Zinn, R. 1985, \apj , 293, 424

\end{thebibliography}
\end{document}